\documentclass[twocolumn,prd,,showpacs,preprintnumbers,amsmath,amssymb,unsortedaddress,epsfig,floatfix]{revtex4}
\usepackage{multirow}
\usepackage{graphicx}
\usepackage{dcolumn}
\usepackage{amsmath}
\usepackage{amssymb}
\usepackage{pslatex}
\usepackage{url}
\usepackage[dvips,hypertex,bookmarks]{hyperref}

\newcommand{\beq}{\begin{equation}}
\newcommand{\eeq}{\end{equation}}

\newcommand{\comment}[1]{}

\begin{document}




\title{LUNASKA experiments using the Australia Telescope Compact Array to search for ultra-high energy neutrinos and develop technology for the lunar Cherenkov technique}

\author{
C.~W.~James$^{1,*}$,
R.~D.~Ekers$^{2}$,
J.~Alvarez-Mu\~{n}iz,$^{3}$,
J.~D.~Bray$^{1,2}$,
R.~A.~McFadden$^{4,2}$, 
C.~J.~Phillips$^{2}$, 
R.~J.~Protheroe$^{1}$,
P.~Roberts$^{2}$
}
\vspace{2mm}
\noindent
\affiliation{
$^{1}$School of Chemistry \& Physics,  Univ.\ of Adelaide, Australia.\\  
$^2$Australia Telescope National Facility, Epping, Australia.\\
$^3$Dept.\ Fisica de Particulas \& IGFAE, Univ. Santiago de Compostela, Spain. \\
$^4$School of Physics, Univ.\ of Melbourne, Australia.\\
$^*$Present address: IMAPP, Radboud Universiteit Nijmegen, The Netherlands.
}
\collaboration{LUNASKA Collaboration}
\noaffiliation

\begin{abstract}
We describe the design, performance, sensitivity and results of our
recent experiments using the Australia Telescope Compact Array (ATCA)
for lunar Cherenkov observations with a very wide ($600$~MHz) bandwidth
and nanosecond timing, including a limit on an
isotropic neutrino flux. We also make a first estimate of the 
effects of small-scale surface roughness on the effective experimental
aperture, finding that contrary to expectations, such roughness will
act to increase the detectability of near-surface events over
the neutrino energy-range at which our experiment is most sensitive
(though distortions to the time-domain pulse profile
may make identification more difficult).
The aim of our ``Lunar UHE Neutrino Astrophysics using
the Square Kilometer Array'' (LUNASKA) project is to develop the
lunar Cherenkov technique of using terrestrial radio telescope arrays for
ultra-high energy (UHE) cosmic ray (CR) and neutrino detection,
and in particular to prepare for using the Square Kilometer Array (SKA) and
its path-finders such as the Australian SKA Pathfinder (ASKAP) and 
the Low Frequency Array (LOFAR) for lunar Cherenkov experiments.  

\end{abstract}



















\maketitle


\section{Introduction}

The origin of the ultra-high energy cosmic rays (UHE CR) ---
protons and atomic nuclei with observed energies above
$10^{18}$~eV and up to at least $2\times 10^{20}$~eV --- is
obscured due to their deflection and scattering in
cosmic magnetic fields.  This makes the flux of all but the
highest energy CR appear almost isotropic with respect to the Galaxy
regardless of their source, so that measurements of arrival
directions cannot reliably be used for source identification. At
the highest energies, the deflection is less, and this allows the
possibility of `seeing' nearby UHE CR sources.  Arrival
directions of UHE CR detected by the Pierre Auger experiment
above $5.6\times 10^{19}$~eV are found to be statistically
correlated with positions of nearby active galactic nuclei (AGN),
which are in turn representative of the large-scale distribution
of matter in the local universe \cite{AugerScience07}. However,
the flux is extremely low, and so the nature of the sources of
UHE CR within this distribution remains at present unresolved.

An alternative means of exploring the origin of UHE CR is to
search for UHE neutrinos.  As first noted by Greisen
\cite{Greisen} and by Zatsepin \& Kuzmin \cite{Zatsepin_Kuzmin},
cosmic rays of sufficient energy will interact (e.g.\ via pion
photo-production) with photons of the 2.725~K cosmic microwave
background (CMB), with the resulting energy-loss
producing a cut-off in the spectrum (the `GZK cut-off') at around $\sim10^{20}$~eV from a
distant source. These same
interactions produce neutrinos from the decay of unstable
secondaries. Several experiments
\cite{Takeda03,Bird95,Connolly06,Abu-Zayyad05, AugerSpectrum08, HiResSpectrum0809} have reported UHE
CR events with energies above $10^{20}$~eV, and therefore a flux
of these ``cosmogenic neutrinos'' is
almost guaranteed.

Significant information on the CR spectrum at the sources is
expected to be preserved in the spectrum of astrophysical neutrinos
\cite{Protheroe04} which varies significantly between different
scenarios of UHE CR production. These include acceleration in the
giant radio lobes of AGN, the decay of super-massive dark matter
particles or topological defects, and $Z$-burst scenarios, the
last of which have already been ruled out by limits placed on an
isotropic flux of UHE neutrinos \cite{Gorham04,Barwick06}. Of
course, neutrinos are not deflected by magnetic fields, and so
should point back to where they were produced, with even a single
detection allowing the possibility of identifying the source of
UHE CR. Here we emphasise that in all models of UHE CR origin we
expect a flux of UHE neutrinos. See
refs.~\cite{ProtheroeClay2004, Falcke2004_SKAscienceCase} for
recent reviews of UHE CR production scenarios and radio
techniques for high-energy cosmic ray and neutrino astrophysics.

\subsection{The lunar Cherenkov technique}
\label{thelunarcherenkovtechnique}

A high-energy particle interacting in a dense medium will produce a
cascade of secondary particles which develops an excess negative
charge by entrainment of electrons from the surrounding material
and positron annihilation in flight.  The charge excess is
roughly proportional to the number of particles in
electromagnetic cascades, which in turn is roughly proportional
to the energy deposited by the cascade.  Askaryan~\cite{Askaryan}
first noted this effect and predicted coherent Cherenkov emission
in dense dielectric media at radio and microwave frequencies
where the wavelength is comparable to the dimensions of the
shower. At wavelengths comparable to the width of the shower, the
coherent emission is in a narrow cone about the Cherenkov angle
$\theta_C=\cos^{-1} (1/n)$ ($n$ is the refractive index), while for
wavelengths comparable to the shower length, the coherent emission is
nearly isotropic. The Askaryan effect has now been
experimentally confirmed in a variety of media
\cite{Saltzberg_GorhamSAND01, Saltzberg_GorhamSALT05,
Saltzberg_GorhamICE07}, with measurements of the radiated
spectrum agreeing with theoretical predictions (e.g.\ ref.\
\cite{Alvarez-Muniz02}). If the interaction medium is transparent to
radio waves, the radiation can readily escape from the medium and
be detected remotely. Since the power in coherent Cherenkov
emission is proportional to the square of the charge excess,
i.e.\ to the square of the energy deposited, extremely high
energy showers should be detectable at very large distances.

The lunar Cherenkov technique, first proposed by Dagkesamanskii
and Zheleznykh~\cite{Dagkesamanskii}, aims to utilize the outer
layers of the Moon (nominally the regolith, a sandy layer of
ejecta covering the Moon to a depth of $\sim$10~m) as a suitable
medium to observe the Askaryan effect. Since the regolith is
transparent at radio frequencies, coherent Cherenkov emission
from sufficiently high-energy particle interactions
(specifically, from UHE cosmic ray and neutrino interactions) in
the regolith should be detectable by Earth-based
radio-telescopes
(Askaryan's original idea was to place detectors on the lunar surface itself).
First attempted by Hankins, Ekers \&
O'Sullivan~\cite{Hankins96,James07} using the Parkes radio
telescope, the Goldstone Lunar UHE neutrino Experiment (GLUE) at
the Goldstone Deep Space Communications Complex \cite{Gorham04},
the experiment at Kalyazin \cite{Beresnyak05}, and the NuMoon
\cite{NuMoon20hARENA2008} experiment at the Westerbork Synthesis
Radio Telescope (WSRT) have subsequently placed limits on an
isotropic flux of UHE neutrinos. Our own project, LUNASKA, aims
to develop the lunar Cherenkov technique for future use of the
SKA.  Observations continue at both WSRT \cite{Scholten06} and
with our own project using the ATCA and using the Parkes radio
telescope (to be reported elsewhere).  The technique has been the
subject of several theoretical and Monte Carlo studies
\cite{ZasHalzenStanev92,Alvarez-Muniz02,GorhamRADHEP01,Beresnyak04,GayleyMutelJaeger2009}
together with our own recent work
\cite{James07,JamesProtheroeLCT109,JamesProtheroe_DirectionalAperture2009}.

Future radio instruments will provide large aperture array (AA)
tile clusters and arrays of small dishes with very broad
bandwidths, with both factors allowing very strong discrimination
against terrestrial radio frequency interference (RFI). The
culmination of the next generation of radio instruments will be
the Square Kilometre Array, to be completed around 2020, with
smaller path-finders such as ASKAP (Australian SKA Pathfinder
\cite{ASKAP07}) to be built in the intervening period. In the
meantime, we have been performing a series of experiments with
the Australia Telescope Compact Array \cite{ATCA}, an array of
six 22~m dishes which were undergoing an upgrade to an eventual
2~GHz bandwidth at the time of the observations
described here. Lunar Cherenkov experiments with these
instruments, together with those proposed for LOFAR
\cite{Scholten06}, represent the foreseeable future of the
technique.  We emphasize that the lunar Cherenkov technique is
very different to conventional radio astronomy and requires
non-standard hardware and signal processing as it is necessary to
detect nanosecond-duration lunar Cherenkov radio pulses coming
from a region too large (the apparent diameter of the Moon is $0.5^{\circ}$)
to image in conventional ways at
nanosecond time resolution. Such
pulses suffer dispersion in the Earth's ionosphere, and our
experiment is the first to correct for this in real-time.

In the present paper we describe our recent experiments using the
ATCA using the lunar Cherenkov technique.  We start by giving an
overview of the experiment, the part of the moon targeted and
observing times which were chosen in order to observe the
Galactic Center and Centaurus A (UHE neutrino flux limits for
these sources are reported in ref.~\cite{ATCA2008data_CenA_GC_limit}), the antennas used,
specialized hardware, triggering and signal processing.  We then
discuss the effects of dead-time, our finite sampling rate and our
approximate de-dispersion on the detection efficiency and
effective observation time.  Finally, we present a new limit to
an isotropic UHE neutrino flux, and discuss why it is important
despite better limits existing from the Antarctic Impulsive
Transient Antenna (ANITA) and the Radio Ice Cherenkov Experiment
(RICE).  In appendices, we outline the effects of the two major
theoretical uncertainties in our aperture and limit calculation
being the UHE neutrino cross-section, and the small-scale lunar
surface roughness. In the latter case, a new approximate
treatment is described, which we use as well as the standard
calculation methods to determine the experimental apertures and
limits given in the main body of the paper.

\section{Description of the Experiment}
\label{ATCA_description}

\begin{figure*}
\centering
\includegraphics[width=0.6 \textwidth]{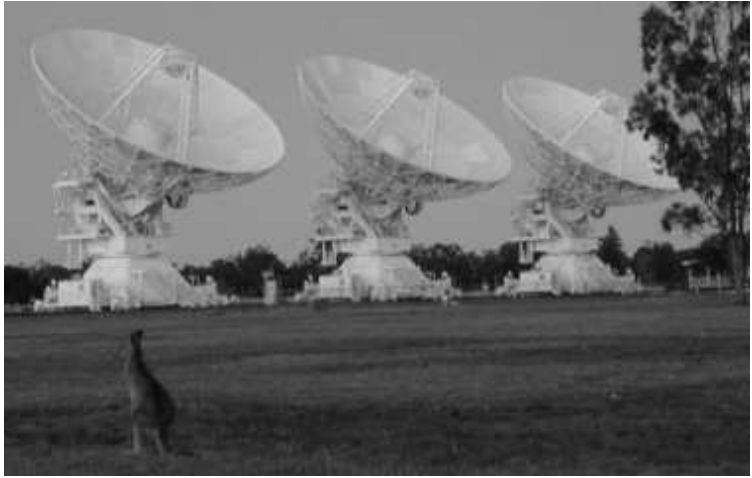}
\caption[Picture of the ATCA]{A photograph of three of the six ATCA antennas. For the observations described here,
the distance between the antennas was greater.}
\label{ATCApicture}
\end{figure*}

The Australia Telescope Compact Array (ATCA -- see Fig.\ \ref{ATCApicture}) is an aperture synthesis
telescope located at latitude -30$^\circ$ near Narrabri, NSW,
Australia.  It consists of six identical $22$~m dishes. One
antenna is fixed in position, while the other five can be moved along a $3$~km
East-West baseline, and also a $\sim 150$~m North-South baseline.
We used three of the moveable dishes on the East-West baseline
for our observations and used triple coincidences with the correct 
relative timing to identify pulses coming from the direction of the Moon

The ATCA was chosen as an ideal SKA test-bed for the lunar
Cherenkov technique because: the antennas have a size comparable
to that expected for the SKA; the beam-size matches the lunar
disk at 1~GHz; it provided us with $600$~MHz of
bandwidth (to be upgraded to 2~GHz); because it is an array
it provides strong timing discrimination against terrestrial RFI;
and because it can give a large aperture sensitivity while seeing the entire
moon \cite{Ekers08}.  Full details of the experiment and
observations are given by James \cite{James_PhD}.

\subsection{Observations}
\label{ATCAobservations}

The observations described here cover two observing periods:
February and May 2008, all at the ATCA. Table \ref{ATCAantennas}
specifies the array configuration used in each observation
period. The baselines used were a compromise between baselines
long enough to resolve correlated thermal emission from the Moon,
and short baselines to reduce the search window in the time
domain for pulses coming from any location on the Moon.  For
these initial observations we implemented our special hardware
on only three of the ATCA antennas: CA01, CA03 and CA05.  

\begin{table*}
\caption[ATCA antenna configurations.]{Positions of the antennas used during the LUNASKA lunar observations, and the baselines. \label{ATCAantennas}}
\begin{center}
\renewcommand{\arraystretch}{1.1}
\begin{tabular}{l l r  c c c r  c c c}
\hline
\hline
{Date} &   Configuration & \multicolumn{3}{c}{ Antenna stations } &  \multicolumn{3}{c}{Baselines (m)}\\
 & & CA01 & CA03 & CA05 & 1-3 & 1-5 & 3-5\\
\hline
Feb 26--28 2008 	& 750B &	W98 & W113 & W148 & 230 & 766 & 536\\
May 18 2008 	& 750A &	W147 & W172 & W195 & 383 & 735 & 352\\ 
May 19 2008 		& EW352 &	W102 & W109 & W125 & 107 & 352 & 245	\\
\hline
\hline
\end{tabular}
\end{center}
\end{table*}

\subsubsection{Observation times and antenna pointings}
\label{ATCAobs_times_pointings}

We had two main considerations in choosing our observation times,
the first being to confine the observations to within the
approximate hours of $10$~pm to $6$~am, in order to have a stable
ionosphere (see Sec.\ \ref{ATCAdedispersion_filters}). The
requirement of the Moon being visible meant that this gave us a
window of perhaps five days in every $29.5$-day synodic lunar
month where the Moon would be sufficiently visible during this
period to warrant observations. The second requirement was that
the Moon be within $\sim 35^{\circ}$ of particular regions of sky
of interest \cite{JamesProtheroe_DirectionalAperture2009}. This
occurs once per $27.3$-day lunar orbit for any given region, so that combined, we
typically had three good and two marginal periods of a few nights
each year in which to observe any given source.

The February 2008 run was tailored to `target' a broad ($\gtrsim
20^{\circ}$) region of the sky near the Galactic Center,
harbouring the closest super-massive black hole to Earth, and a
potential accelerator of UHE CR. The Galactic Center may also be
a source of UHE CR and neutrinos through the decay of massive
particles in its dark matter halo (see
\cite{AloisioTortorici2008} and references therein).  Preliminary
calculations showed that for beam-sizes similar to that of the
ATCA, the greatest total effective aperture (and hence
sensitivity to an isotropic or very broadly-distributed flux) is
achieved when pointing the antennas at the center of the
Moon, so all the limb is at approximately the half power point
of the antenna beam. Since any UHE neutrino flux from this region is likely to
be broadly-distributed, we used this pointing for these runs. Our
May 2008 observing period targeted Centaurus A only, the nearest
active galaxy which could potentially account for some of the UHE
CR events observed by the Pierre Auger observatory
\cite{AugerScience07}, and to achieve the maximum sensitivity to
UHE neutrinos from this source we pointed towards the portion of the
lunar limb closest to Centaurus A so it is observed at full sensitivity.

\subsection{Specialized hardware}

\begin{figure*}
\centering
\includegraphics[width=0.6 \textwidth, angle=270]{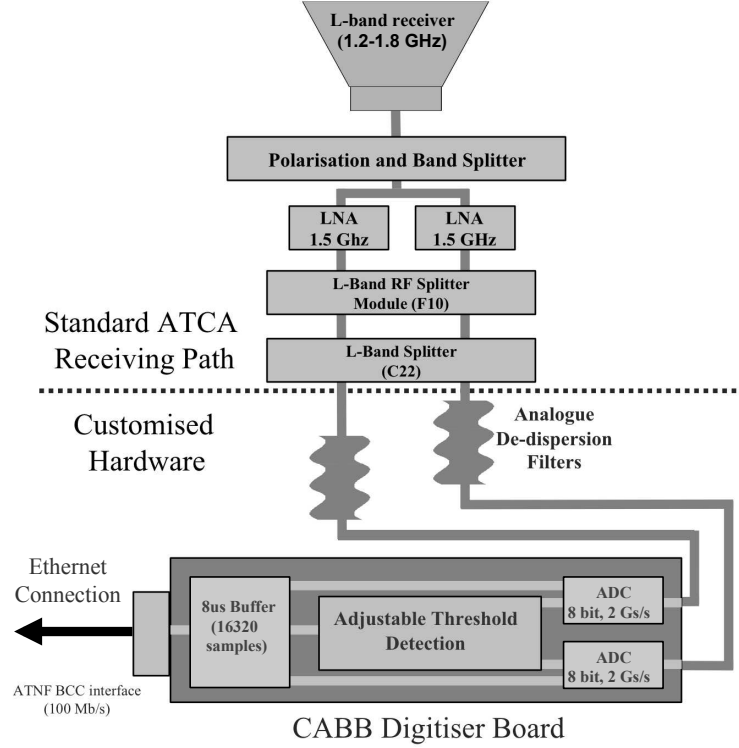}
\caption[Diagram of the signal path at each antenna.]{Diagram of the signal path at each antenna. Abbreviations are: `LNA': low noise amplifier, `ADC': analogue--digital converter.}
\label{ATCAblock_diagram}
\end{figure*}

The background signal above which any genuine nanosecond-duration lunar
Cherenkov pulses had to be detected consisted of two components: random noise fluctuations,
mainly thermal emission from the lunar disk and to a lesser extent system noise and
Galactic plane synchrotron emission; and man-made RFI. Fig.\
\ref{ATCAblock_diagram} gives a diagram of the hardware and
signal path at each antenna. In order to perform a search for
short-duration lunar pulses, against a background of
thermal noise fluctuations and RFI, we had to build specialized
hardware to detect and store candidate events in real time. For
this we used the digital, field-programmable gate array (FPGA)
based analog-to-digital converters (ADC) used in the Compact
Array Broad Band (CABB) upgrade \cite{CABB_upgrade}, each of
which could digitize and perform simple logic on two data streams
at a rate of $2.048$~GHz. As well as this, we required
specialized software running on the control-room computers to
interface with the CABB analog-to-digital converter boards, and a hardware method to
correct for the dispersive effects of the Earth's ionosphere.

\subsubsection{Data channels and signal path}

During each observation period, the received signal -- 
split into two orthogonal linear polarizations, which we
arbitrarily label `A' and `B' --
was processed by our specialized hardware at each
of three antennas, providing a total of six output data streams. 
By connecting to a monitor point at the receivers in each antenna
we bypassed the normal narrow-band ATCA intermediate frequency (IF) system and obtained a nominal
bandwidth of $600$~MHz between $1.2$ and $1.8$~GHz. For May
$2008$ a high-pass filter was added to remove a strong $0.9$~GHz (aliased to
$1.1$~GHz) narrow-bandwidth RFI from a known transmitter which
was perturbing our detection threshold.

Each polarized data stream was fed through an analog
de-dispersion filter before being sampled by a CABB analog--digital converter
board. This operated at $2.048$~Gs/s with $8$-bit effective
precision.
Since both the 100 Mb/s connection from each antenna
to the central control room (where all antenna signals could
be combined) and the CABB correlator architecture were
inadequate to handle the full raw data rate of
$2 \times 8 \times 2048$ megabit/s per antenna,
we had to reduce our data volume by triggering
independently at each antenna and returning only blocks of data
containing candidate events.
The signal was
copied into both a running buffer and passed to a real-time
trigger algorithm, and on fulfilling the trigger conditions a
portion of the buffer was returned to the control room and
recorded. During this recording process, the buffer was unable to
respond to further triggers, and the experiment was temporarily
blind to any events.  The period of this dead-time depended on
the length of the buffer to be returned. 

\subsubsection{Trigger logic and levels}
\label{ATCAtriggering}

The trigger algorithm was set up to be a simple threshold trigger
at each antenna -- if the square of any single sample on either
the A or B polarization data streams was above a certain value,
both polarizations were returned.  The thresholds were adjusted
occasionally to keep the trigger rates on each receiver output
constant at approximately 40-50 Hz corresponding to $\sim
5\sigma$, where we use $\sigma$ as short-hand for the rms voltage
$V_{\rm rms}$ in the output channel. Even with our 8-bit
sampling we were barely able to adjust these threshold with
sufficient precision ($< 0.1 \sigma$ increments) while
maintaining a reasonable dynamic range for any detected event.
The gain was adjusted to give an RMS sampler output of approximately 10 ADC
digitization units (a.d.u.) and hence a maximum of 12.8$\sigma$ (128 a.d.u) before saturation for an 8-bit signal.


\subsection{Dead-time and efficiency}
\label{ATCAefficiency}

A certain degree of dead-time loss is suffered for every
trigger. As all three antennas need to be `on' to record an
event, it is important to avoid setting the thresholds too low
(trigger rates too high) as this can make the effective
observation time negligible. This dead-time can be easily
measured by setting the thresholds to zero and recording the
maximum trigger rate for a given buffer length. Such measurements
were performed at each observation period, and the results are
recorded in Table \ref{ATCArates_table}. We see that for
a buffer length of $256$ samples the dead-time per trigger 
was approximately $1$~ms. 

\begin{figure*}
\begin{center}
\includegraphics[width=\textwidth]{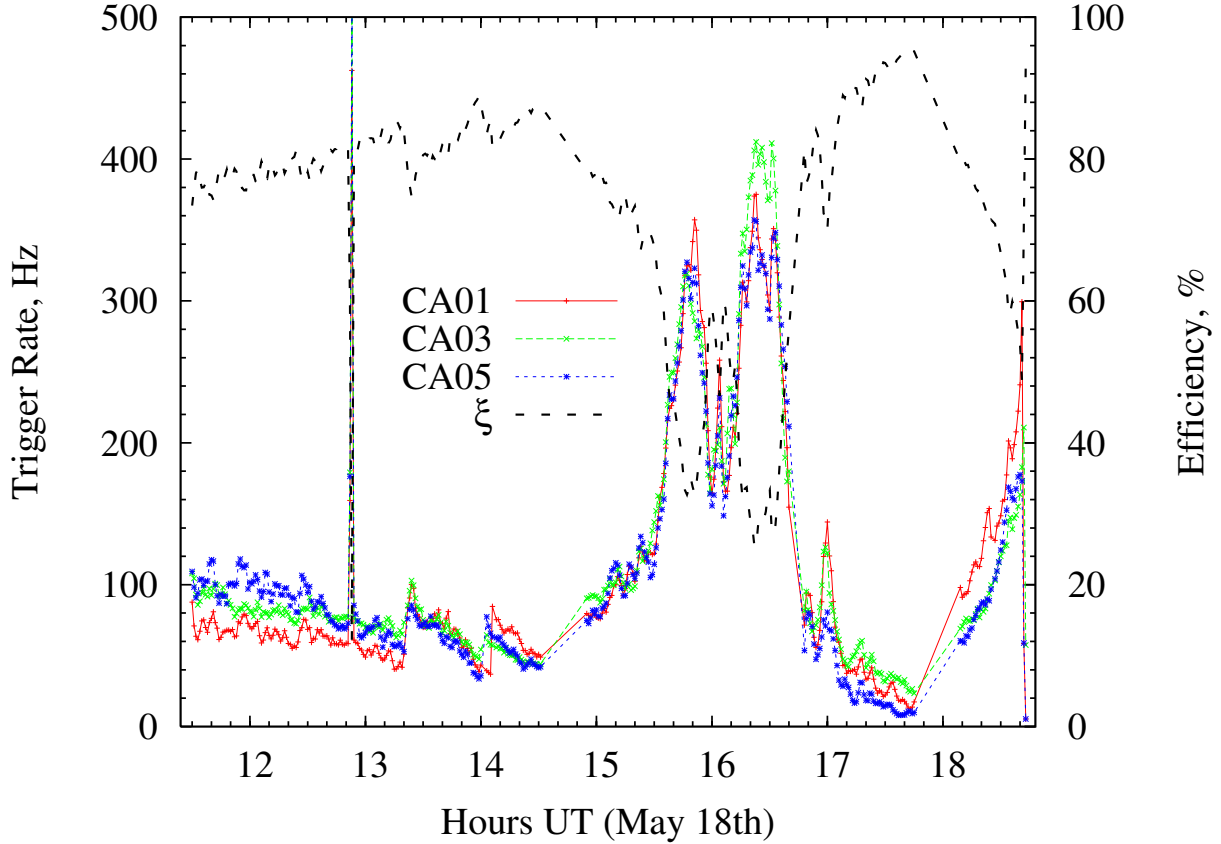}
\caption{(color online) Trigger rates for the three antennas (three lower curves) and efficiency (upper curve) from Eq.\ (\ref{ATCAtrig_rate_1}) for the night of May $18^{\rm th}$, 2008. The increased trigger rate at UT 15:00-17:00 results from a `Type 1' noise feature caused by an unknown source -- see Sec.\ \ref{RFI} and the top-centre of Fig.\ \ref{ATCAtriple_ATCAevents}.}
\label{ATCAratesandsuch}
\end{center}
\end{figure*}

The
efficiency of the experiment can be defined as the time-fraction
when all three antennas are sampling and ready to trigger. For a
sampling rate $r_i$ (Hz) on antenna $i$, maximum rate $R_i$, and
purely random trigger events, the efficiency $\xi$ is given by:
\begin{eqnarray}
\xi & = & \Pi_i \left(1-\frac{r_i}{R_i}\right) \label{ATCAtrig_rate_1}
\end{eqnarray}
where the $i$ multiplies over all three antennas.  In Fig.\
\ref{ATCAratesandsuch} we plot the trigger rates for all antennas
for the 18 May 2008 observations and the efficiency $\xi$
calculated as in Eq.~\ref{ATCAtrig_rate_1}, assuming a constant
$R_i=1040$ from Table \ref{ATCArates_table} corresponding to our
buffer length of $256$ samples.  A short-duration burst of RFI is
evident at UT 12:50 in Fig.\ \ref{ATCAratesandsuch}, as is
a large increase in the background between UT 15:00 and UT 17:00.
During these periods of intense RFI the efficiency (upper dashed line)
is significantly reduced.  The effective observation time $t_{\rm
eff}$ was determined by integrating the efficiency over the
observation time $t_{\rm obs}$, and this is given in Table
\ref{ATCAtefftable} together with the average efficiency
$\bar{\xi}=t_{\rm eff}/t_{\rm obs}$.

\begin{table*}
\begin{center}
\renewcommand{\arraystretch}{1.3}
\caption{Maximum trigger rates (Hz) as a function of buffer length for 18 May $2008$.} 
\label{ATCArates_table}
\begin{tabular}{l ccccccccc }
\hline
\hline
Buffer length & 16256 & 8192 & 4096 & 2048 & 1028 & 512 & 256 & 128 & 64 \\
\hline
Trigger rate & 22 & 42 & 83 & 163.5 & 317.5 & 581 & 1040 & 1690 & 2450 \\
\hline
\hline
\end{tabular}
\end{center}
\end{table*}

\begin{table}
\caption[Efficiency of the 2008 observations.]{Raw observation time $t_{\rm obs}$ (minutes), mean efficiency $\bar{\xi}$ (\%), and effective observation time $t_{\rm eff}$ (minutes), for all observation periods.}
\begin{center}
\renewcommand{\arraystretch}{1.4}
\begin{tabular}{l c c c }
\hline
\hline 
{Date}	& ~  $t_{\rm obs}$ & ~  ${\bar{\xi}}$	& ~  $t_{\rm eff}$ \\
\hline 
26 {Feb.\ 2008}	& ~  239	& ~  86	& ~  204 \\
27 {Feb.\ 2008}	& ~  319	& ~  87	& ~  277 \\
28 {Feb.\ 2008}	& ~  314	& ~  87	& ~  274 \\
17 {May\ 2008}	& ~  324	& ~  69	& ~  224 \\
18 {May\ 2008}	& ~  376	& ~  73	& ~  274 \\
19 {May\ 2008}	& ~  440	& ~  72	& ~  316 \\
\hline
\hline
\end{tabular} \label{ATCAtefftable}
\end{center}
\end{table}

\subsection{Ionospheric dispersion}

For our experimental bandwidth of $600$~MHz centered at $1.5$~GHz, the effects of dispersion in the Earth's ionosphere are significant. The dispersion is due to a frequency-dependent refractive index caused by free (ionized) electrons in the ionosphere. Using the standard measure for the number of electrons (total electron content units, or TECU: $10^{16}~e^-$/cm$^2$), the time-delay $\delta t$ relative to a vacuum for a frequency $\nu$ is given by Eq.\ (\ref{ATCAdelay_eqn}):
\begin{eqnarray}
\delta t & = & 1.34~10^{-7} \, {\rm TECU} \, \nu^{-2} \label{ATCAdelay_eqn}.
\end{eqnarray}
Of more use is the dispersion $\Delta t$ over a bandwidth $\Delta \nu$, given by 
\begin{eqnarray}
\Delta t & = & 1.34~10^{-7} \, {\rm TECU} \, \left(\nu_{\rm min}^{-2}-\nu_{\rm max}^{-2} \right) \label{ATCAdispersion_eqn} \\
\Delta t & \approx & 2.68~10^{-7} \, {\rm TECU} \, \Delta \nu  \, \bar{\nu}^{-3} \label{ATCAapprox_dispersion_eqn} 
\end{eqnarray}
if $\Delta \nu \ll \bar{\nu}$, where $\bar{\nu}$ is the mean frequency $(\nu_{\rm min} + \nu_{\rm max})/2$. Note that in terms of phase delay, the correction goes as $\bar{\nu}^{-2}$.

\subsubsection{De-dispersion filters}
\label{ATCAdedispersion_filters}

\begin{figure}
\begin{center}
\includegraphics[width=0.49\textwidth]{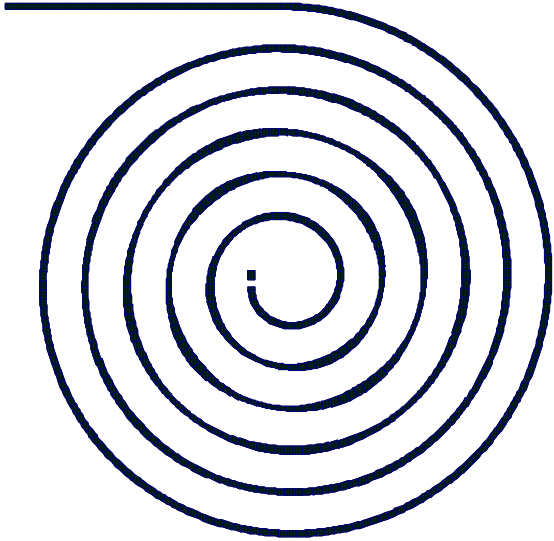}
\caption{Printed circuit board layout of analog de-dispersion filter.  Note the variation in the width of the waveguide along its length. The physical size is approximately $25$x$25$~cm.
\label{FilterPhoto}}
\end{center}
\end{figure}

Implementing a digital de-dispersion filter running at this speed
was too difficult at the time, so we used analog de-dispersion filters
designed by Roberts \cite{PaulRobers08}. Each filter was a variable-width
waveguide of approximately one meter in length constructed as a
spiral for compactness (Fig.~\ref{FilterPhoto}), with the output
being the continuous sum of reflections along the length. Upon
reflection, high frequencies experienced a greater delay than
low frequencies, with the design such that this cancelled out the
delay due to ionospheric dispersion at low frequencies. Thus an
in-phase signal (e.g.\ coherent Cherenkov radiation from a UHE
particle interaction in the Moon) entering the top of the
ionosphere should appear in-phase after de-dispersion, provided
the correct dispersion measure was used.

\begin{figure}
\begin{center}
\includegraphics[width=0.49\textwidth]{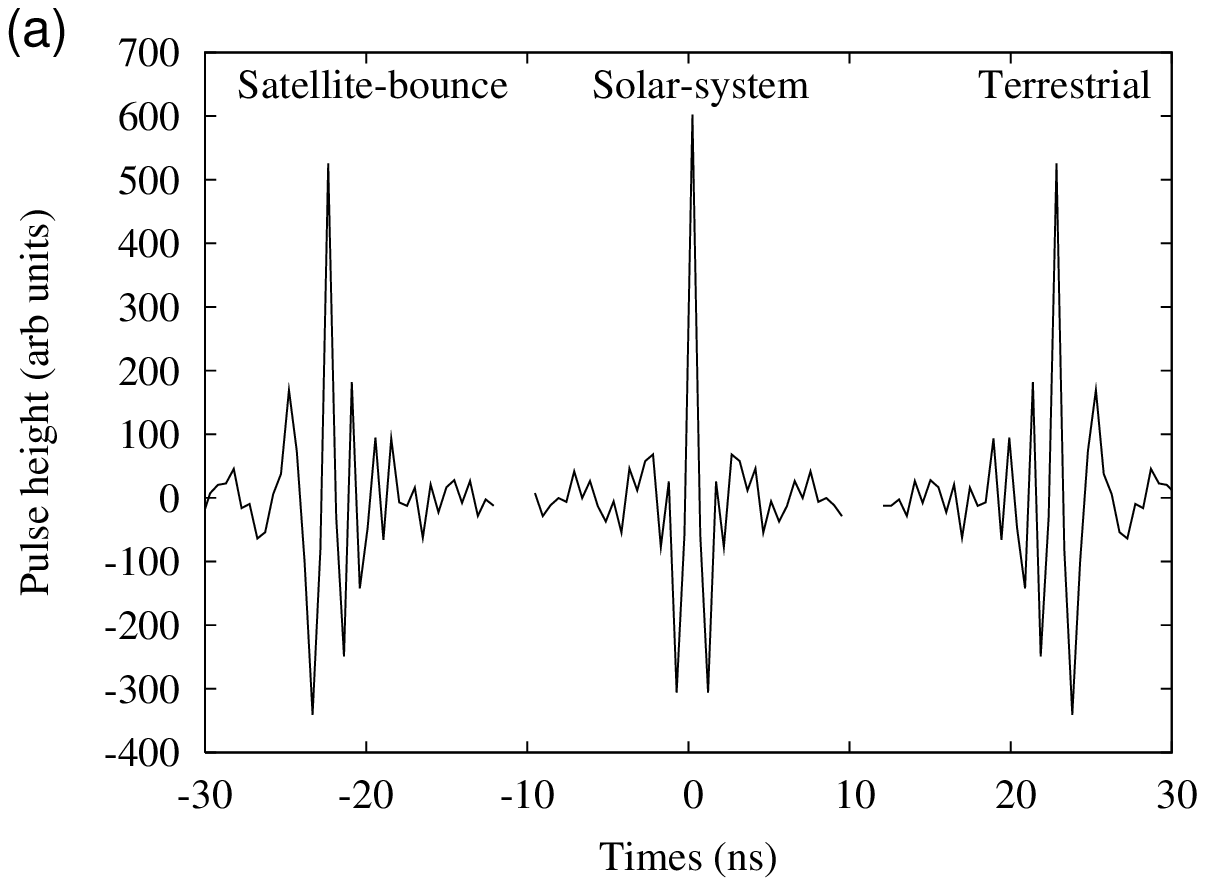}
\includegraphics[width=0.49\textwidth]{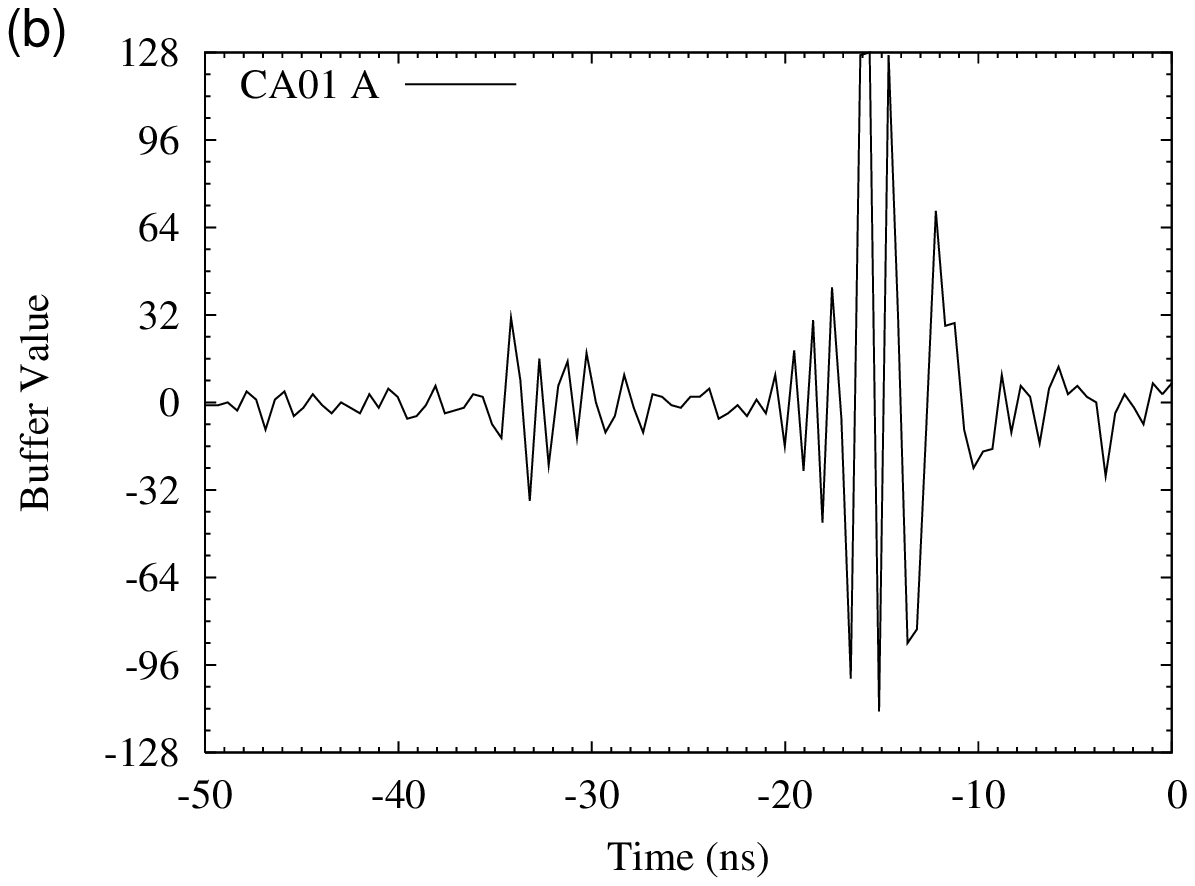}
\caption[Observed noise-calibration pulse and origin-dependent predictions.]{(a) Predictions of an observed impulse of different origins.  (b) A typical noise-calibration pulse -- in this case, from CA01 A on Feb $26^{\rm th}$. The small pre-pulse at $-35$~ns is caused by a spurious reflection from the filter connection point.
\label{ATCAnoise_cal_pulse}}
\end{center}
\end{figure}

We used NASA data \cite{NASA_crustal} to predict typical values
of the ionospheric dispersion at Narrabri, with results reported
in Ref.\ \cite{McFadden07}.  Since we were still near solar
minimum these results showed that the ionosphere over Narrabri was
comparatively stable between the hours of $10$~pm and $6$~am,
with low and predictable vertical total electron content (VTEC)
measures around $7 \pm 1.3$~TECU. This corresponds to a differential
vertical delay of $3.6$~ns over a $1.2$-$1.8$~GHz
bandwidth. Since the actual delay will depend on the slant
angle, we chose to build the filters assuming a $5$~ns delay over
the band, i.e.\ a TEC along the line of sight (slant TEC, or
STEC) of $10$~TECU; this is also equivalent to the mean VTEC of
$7$~TECU with a lunar elevation of $47^{\circ}$. Therefore we
expected to lose some sensitivity when the Moon was directly
overhead, and also very near the horizon. The sensitivity lost
due to deviations of the actual VTEC from the mean and variations
in lunar elevation is discussed in Sec.\
\ref{ATCAeffects_of_dedispersion}.  Fig.\
\ref{ATCAnoise_cal_pulse}(a) shows the expected effect of
de-dispersion on pulses of different origin: satellite bounce (of
RFI), solar system (lunar Cherenkov) and terrestrial (RFI).
These have been modeled respectively as a flat frequency spectrum
dispersed once (dispersed twice then de-dispersed once), a
$\nu^2$ spectrum expected for coherent Cherenkov emission
unchanged (dispersed once then de-dispersed once), and a flat
frequency spectrum de-dispersed once.  In all cases the Fourier
inverse has been taken over 1.2-1.8 GHz, i.e.\ the pulse has been
band-limited to 1.2-1.8 GHz.

\subsection{The noise diode}
\label{ATCAnoise_diode}

A noise-calibration diode is located in the receiver and used to calibrate
the power through measuring the system temperature, $T_{\rm sys}$,
during normal ATCA observations.
During testing, we observed unexpected strong pulses at a rate of
approximately 8 Hz (Fig.\ \ref{ATCAnoise_cal_pulse}(b)). It was
discovered that these pulses were generated by the switching of
the noise-calibration diode. Note that, as expected, the pulse in
Fig.~\ref{ATCAnoise_cal_pulse}(b) closely resembles that expected
for a terrestrial source.

Since the noise diode could be turned
on and off from the control room, it could be used to generate
approximately coincident false triggers between the three
antennas. The time difference in the noise diode switching between
antennas was small ($<1$~$\mu$s) and the scatter about this
offset was $\sim200$~ns. Using this simple procedure to generate
approximately coincident false triggers became a useful part of our
experimental procedure as discussed in Section IV-A.


\section{Sampling and  Dispersive Effects on Sensitivity}
\label{ATCAeffects_of_dedispersion}

The effects of loss of sensitivity due to a finite sampling rate
(compared to an infinite sampling rate) and
de-dispersion depend on the frequency spectrum of the radio pulse
and the bandwidth of the detector.  The expected lunar Cherenkov
pulse spectrum depends on many factors, including lunar surface
roughness and orientation, the dimensions and direction of the
electromagnetic cascade in the lunar regolith, and on the
neutrino energy -- these are discussed in detail elsewhere
\cite{JamesProtheroeLCT109}.  The range of possible spectra is
very broad. However, for our purposes we consider two extreme
cases.  We are concerned with the relative strengths of the
high-frequency and low-frequency components. Near the minimum
detectable cascade energy for our experiment ($10^{20}$~eV --
see Sec.\ \ref{ATCA_isotropic_app}),
only fully coherent emission will be detectable. The electric
field spectrum (V/m/MHz) will therefore have the form
$E(\nu) = A \nu$, which gives the greatest possible weight to
the high-frequency component.

The other extreme is given by a high-energy shower,
at shallow depth and viewed at a large
angle away from the Cherenkov angle where the emission from the
cascade is becoming incoherent. In this case we expect $E(\nu) =
B \nu \exp(-C \nu^2)$ based on approximate fits to the spectrum
far from the Cherenkov angle for hadronic showers
\cite{Alvarez-Muniz98}.  For any observed pulse of given total
power we do not know the shape of the spectrum to expect if the
pulse is indeed of lunar origin, and so we consider the two
extreme possibilities.  For our `high energy' we take $10^{23}$~eV,
since above this range strong limits on a neutrino flux from the
NuMoon \cite{NuMoon20hARENA2008} and FORTE \cite{Lehtinen04} experiments
made a detection extremely unlikely. For the same reasons, $10^{23}$~eV
is also the most energetic neutrinos to which we simulate our effective
apertures and limit, which can at most produce cascades of energy $10^{23}$~eV.
Setting the peak power in the bandwidth
to be the same for the two extreme cases, corresponding to
$10^{20}$~eV and $10^{23}$~eV neutrinos, and taking $B/ A =
(10^{23}/10^{20})$ allows the constant $C$ to be found as
follows.  Since the electric field of a pulse (wave packet) at
the antenna may be written as
\begin{eqnarray}
E(t) = \int_{-\infty}^{\infty} E(\nu) e^{2\pi i \nu (t-t_0)} d\nu,
\end{eqnarray}
and since the power is proportional to $|E(t)|^2$,
the peak power occurs at $t=t_0$ and so  is proportional to
\begin{eqnarray}
|E(t_0)|^2 = \left|\int_{-\infty}^{\infty} E(\nu)  d\nu\right|^2.
\end{eqnarray}
Hence, for equal peak powers for the two extreme forms of possible spectra we may solve 
\begin{eqnarray}
A \int_{\nu_1}^{\nu_2} \nu d \nu & = & B \int_{\nu_1}^{\nu_2} \nu \exp(-C \nu^2) d \nu .
\label{ATCAABCeqn}
\end{eqnarray}
For $B/ A = 10^{3}$, $\nu_1=1.2$~GHz,  and $\nu_2=1.8$~GHz we find $C=3.515$ GHz$^{-2}$. 
We shall use these results when finding the uncertainty in sensitivity due to the unknown spectral shape.

\subsection{The effects of a finite sampling rate only}

\begin{figure}
\begin{center}
\includegraphics[width=0.49\textwidth]{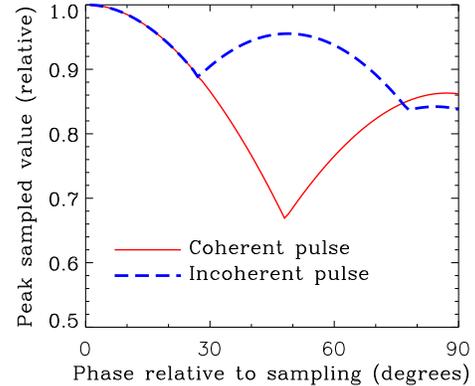}
\caption{(color online) Maximum sampled value as a fraction of
the true pulse height as a function of the arbitrary phase offset
for sampling rate of 2.048~Gs/s and two extreme lunar Cherenkov
pulse types.
\label{ATCAsampling_loss}} 
\end{center}
\end{figure}

Our sampling rate of $2.048$~Gs/s was greater than the Nyquist rate
of $1.2$~Gs/s for our nominal $600$~MHz bandwidth, and allowed
for perfect reconstruction of the signal in the frequency range
$1.024$ to $2.048$~GHz at arbitrary time resolution
(assuming no signals outside this range). However, experiments
(such as this one at the ATCA) which are
pushing the current FPGA limits for high-speed signal processing
can only use simple algorithms based on the pulse height of a single
sample as captured by the sampling threshold. For any finite
sampling rate (including the Nyquist rate), there will be a
random offset in the phase of the ADC digitization times between
the actual peak of the pulse and the sampling times. Fig.\
\ref{ATCAsampling_loss} plots the peak pulse height as a function
of the arbitrary phase offset for the sampling rate of
$2.048$~Gs/s as used in our experiment and the two extreme spectra
of lunar Cherenkov emission.  The peak sampled value is seen to
vary by $\sim$30\% in the case where the Cherenkov emission is
fully coherent, and by $\sim$15\% when the emission is becoming
incoherent.  Hence, with our simple trigger logic, there will be
a triggering inefficiency due to our finite sampling rate causing
the peak sampled voltage occasionally to be less than the trigger
level, even if the actual peak voltage in a pulse was above it.
In future experiments, we will use a more complex trigger algorithm which
works off multiple sampled values in order to reduce this loss, which is 
a more efficient remedy than increasing the sampling rate.

\subsection{Effects due to de-dispersion}

\begin{figure}
\begin{center}
\includegraphics[width=0.49\textwidth]{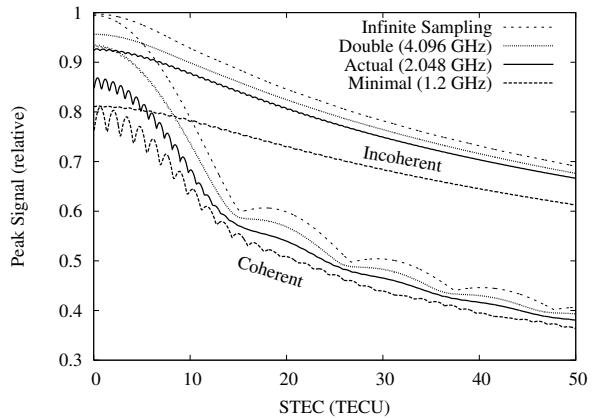}
\caption[Fraction of signal peak detected as a function of
ionospheric dispersion.]{Greatest detected amplitude of the
time-domain signal relative to the maximum undispersed amplitude,
for  two extreme lunar Cherenkov pulse types and four sampling rates, as a function of
the difference between the line-of-sight total electron content,
in TECU, and the constant value used in the de-dispersion.}
\label{ATCAionospheric_loss}
\end{center}
\end{figure}

In order to quantify the effect of using the constant TEC value
built into our de-dispersion filters, we take pulses of the two
extreme types discussed above, disperse them for different
line-of-sight TEC values, de-disperse them using the constant TEC
value built into our de-dispersion filters, and finally simulate
sampling by the ADC.  All possible offsets of the pulse-peak
arrival time with respect to the sampling times, or ``base
phase offsets'', were modeled in this process.  For each
combination of intrinsic spectrum, base phase offset, and
dispersion measure, we calculate the peak pulse strength in the
time domain. Averaging this over all base phase offsets (which
will be random) and dividing by the magnitude of the peak
undispersed pulse at zero phase offset we obtain the peak signal
strength as a function of line-of-sight TEC shown in Fig.\
\ref{ATCAionospheric_loss}.  As well as the sampling rate of 2048
GHz used, we also show results for sampling rates a factor of 2
higher and lower.  The upper and lower sets of lines are for
pulses due to incoherent and coherent Cherenkov emission by the
lunar cascade as a whole, respectively. The pulse for
fully-coherent Cherenkov radiation is most adversely affected by
ionospheric dispersion because the signal is spread over the
largest frequency range. The rapid oscillations in average peak
amplitude with changing line-of-sight TEC for the coherent
Cherenkov pulse is due to the combined effect of the sharp band
edges and the de-dispersion function. The mean values from Fig.\
\ref{ATCAsampling_loss} correspond to zero line-of-sight TEC for
the case of a 2048 GHz sampling rate, allowing a comparison
between the effects of our finite sampling rate and dispersion.  We note
that the loss in sensitivity due to errors caused by using a
constant TEC value for de-dispersion (typically differing from
the true value by less than 4 TECU) are less than the finite
sampling rate errors for our observations using a simple trigger
algorithm.
\subsection{Loss of experimental sensitivity}
\label{ATCA_actual_tec_loss}

\begin{figure*}
\begin{center}
\includegraphics[width=0.7\textwidth]{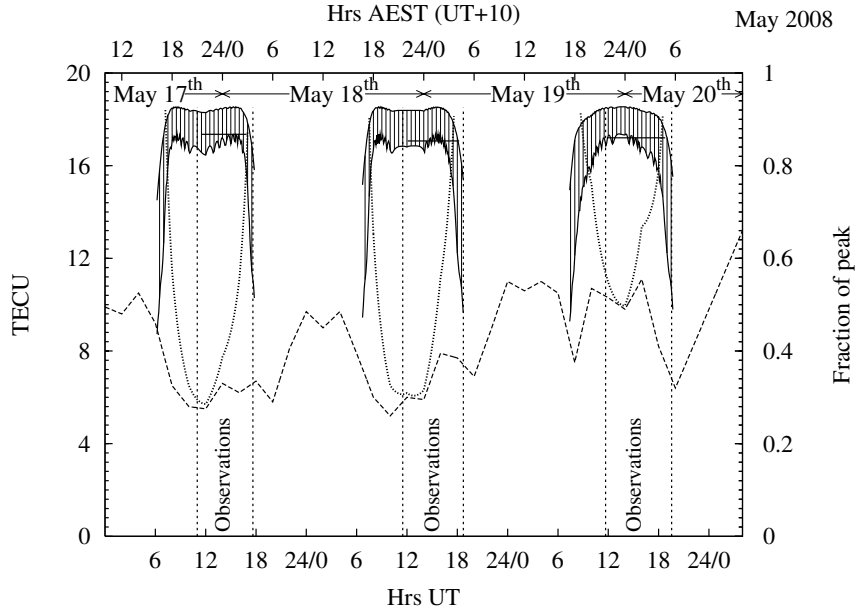}
\caption[Ionospheric influence over our observation
periods.]{Ionospheric influence over our May 2008 observations. Dashed curves:
measured VTEC (TECU) from Ref.\ \cite{NASA_crustal}. Dotted
curves: STEC (slant TEC along the line of sight). Solid curves:
mean fractional loss for coherent (lower curve) and incoherent
(upper curve) lunar Cherenkov pulses due to ionospheric effects and triggering inefficiency
due to our non-infinite sampling rate (see Fig.~\ref{ATCAsampling_loss}): the shading
gives the range. Vertical lines indicate the times of the
observations, horizontal lines give the
fractional loss averaged over each observation and over the two
extreme cases for the expected spectrum.}
\label{ATCAall_tecu_graph}
\end{center}
\end{figure*}

The loss of sensitivity due to sampling and dispersive effects
can be calculated using the measured values of total
electron content (TEC) as a function of
elevation, which were determined after the observations \cite{NASA_crustal}.
The results are given in Fig.\
\ref{ATCAall_tecu_graph}. Using a linear interpolation between
these points and the known lunar elevation gives the
line-of-sight TEC (slant TEC, or STEC) measure (dotted lines).
At low elevations, the line of sight will
probe a large horizontal distance, so using a constant VTEC
measure may not be appropriate. However, since the TEC goes as
$1/\sin(\rm elevation)$ and consequently blows up at low
elevations, the sensitivity in this regime will be low in any
case. Combined with the mean losses for the two spectra in Fig.\
\ref{ATCAionospheric_loss}, the range of losses for the
experimental periods is calculated as the shaded regions in Fig.~\ref{ATCAall_tecu_graph} which are bounded
by the extreme spectra of coherent/incoherent lunar Cherenkov emission by lunar cascades.

The effect of using a fixed STEC value for the de-dispersion is estimated by taking the mean loss over both observation time and spectral type, giving equal weighting to both the coherent and incoherent spectra. The resulting mean detected signal fractions are given in Table \ref{ATCAdispersion_effects}.

\begin{table}
\caption[Estimated fractions of the peak signal detected during ATCA observations.]{Estimated nightly average fractions of the peak signal detected (\%) for the observations periods indicated. The best case corresponds to incoherent Cherenkov emission from lunar cascades, the worst case to completely coherent lunar Cherenkov emission signals, as discussed in text.} \label{ATCAdispersion_effects}
\begin{center}
\renewcommand{\arraystretch}{1.1}
\begin{tabular}{ l c c c  c  c c c  c }
\hline
\hline
Period &  \multicolumn{4}{c }{February 2008} & \multicolumn{4}{c}{May 2008} \\
Date &  $26^{\rm th}$  & $27^{\rm th}$ & $28^{\rm th}$ & Mean & $17^{\rm th}$ & $18^{\rm th}$ & $19^{\rm th}$ & Mean\\
\hline
Best (\%) & 92 & 91 & 91 & 91 & 91 & 91 & 91 & 91 \\
Worst (\%) & 84 & 80 & 81 & 82 & 82 & 80 & 81 & 81\\
Mean (\%) & 88 & 86 & 86 & 87 & 87 & 86 & 86 & 86\\
\hline
\hline
\end{tabular}
\end{center}
\end{table}

\section{Relative Timing Calibration with Astronomical Point Sources and RFI sources}
\label{ATCAtime_alignment}

We had counters recording the number of samples at each antenna, the value of which was returned
with each triggered event. These could be converted to clocks accurate to $\sim0.5$~ns. However, at the time of our experiment these clocks had unknown timing offsets between them, which had to be determined to allow a sufficiently rigorous pulse search. 
To calibrate the times, we required both a common signal in each
antenna with some known time-delay, and a method to trigger the
buffers with sufficient simultaneity that enough of the common
signal seen by all three antennas would be captured to produce a
significant correlation.  We used the noise-calibration pulses
(see Sec.~\ref{ATCAnoise_diode}) as our trigger, and the very
bright discrete radio sources 3C273 and 3C274 (M87) as our
correlated signal.

The QSO 3C273 was chosen since it was the brightest ($47$~Jy at 1.4 GHz)
point-like source near the Moon at the time of the observations, and
would thus give a strong correlation over all baselines. The radio
galaxy M87 is brighter ($215$~Jy at 1.4 GHz) \cite{ATCA} but resolved
on our long baselines, and was chosen to maximise the correlated signal
over our shortest (CA01-CA03) baseline only.

\subsection{Timing observations}

To calibrate the timing, we pointed the antennas at either 3C273
or M87, set the buffer length to maximum, switched the noise-calibration
on, and set the trigger thresholds such that we were triggering only off
the noise-calibration pulses, at roughly $8$~Hz. We typically observed
in this calibration mode for a few minutes at a time and thus took of
order $2000$~pulses, repeating this procedure a few times each night.

\begin{figure}
\begin{center}
\includegraphics[width=0.49\textwidth]{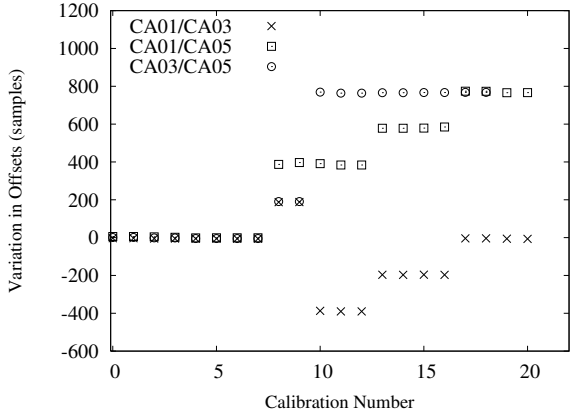}
\caption[Consistency check of correlation times for the 2008 data.]{Consistency check of correlation times in the February  data: alignment relative to the first calibration of the February period, in ($x$-axis) chronological order.
 \label{ATCAalignment_figs}}
\end{center}
\end{figure}

The timing offsets, $\Delta t$, measured in samples, are given in
Fig.\ \ref{ATCAalignment_figs} for the February observation
periods. The vertical axis shows the absolute time offsets
between antennas $j$ and $i$, $\Delta t^0_{ij}$, relative to the
first calibration after each clock reset -- i.e.\ the first data
points have been adjusted to $0$.  This
adjustment is the timing calibration offset.  The
expectation was that all data would therefore have $y$-values
near 0.  Obviously this is not the case, and we see the time
offsets $\Delta t^0_{ij}$ jump around in multiples of $192$
samples ($93.75$~ns).

We eventually discovered that these 192-sample offsets were a
hardware fault triggered whenever we changed the buffer lengths.
In the majority of cases we were able to use our log-book and RFI
in the data to successfully identify when these offsets occurred
and make the appropriate correction.  In a small number of time
blocks we still had ambiguity so we searched for event
coincidences using all possible 192-sample offsets.  Because the
triple coincidence requirement is so strong this did not have an
impact on the final sensitivity.

\section{RFI}
\label{RFI}

\begin{table}
\caption{Number of two-fold and three-fold coincidences, within an $8000$ sample ($\sim 4$~$\mu$s) window for each night (the count of two-fold events also includes the three-fold events). 
\label{ATCAstats_combined_tbl}}
\begin{center}
\renewcommand{\arraystretch}{1.2}
\begin{tabular}{ l r r r r }
\hline
\hline
Date & CA01/03 & CA01/05 & CA03/05 & CA01/03/05 \\
\hline
February    26 &   6445 &   1286 &   1533 &    449 \\
February    27 &  68894 &  39898 &  43224 &  30051 \\
February    28 &  23296 &   8590 &  11072 &   6781 \\
May    17 &   2344 &   1925 &   1994 &     96 \\
May      18 &  21774 &  19445 &  20635 &   3437 \\
May     19 & 114383 &  74311 &  71313 &  57493 \\
\hline
\hline
\end{tabular}
\end{center}
\end{table}

After correcting the timing for the 192-sample jumps,
 we looked at coincident
triggers within the physically possible time range for signals
coming from outside the array.  The number of two-fold and
three-fold coincidences for each observation night are given in
Table~\ref{ATCAstats_combined_tbl}.  
The most obvious result is
the extremely large number of two-fold coincidences, and the
large number of three-fold coincidences compared to two-fold
coincidences. 

The expected rate (Hz) of two-fold coincidences, $R_{ij}$, and three-fold
coincidences, $R_{135}$, from purely random arrival times is given by 
\begin{eqnarray}
R_{ij} & = & R_i R_j W_t \\
R_{135} & = & R_i R_j R_k W_t^2 \\
R_{135} / R_{ij} & = & R_k W_t 
\label{ATCAcoincidence_eqn}.
\end{eqnarray}
where $R_{i}$ is the rate (Hz) of single triggers in antenna $i$
and $W_t$ is the time window (seconds) required for a
coincidence.  Hence, the ratio between three-fold and two-fold
coincidences increases with the trigger rate.  For a maximum
trigger rate $R_i$ of $3$~kHz and time window $W_t=\pm
3.906~10^{-6}$~s ($8000$ samples), the two-fold rate is
$70$~Hz. However, the three-fold rate is only $1.6$~Hz, i.e.\
only $\sim2$\% of the two-fold rate. While at times the ratio of
two-fold to three-fold coincidences matched this expectation
exactly, there were also some periods within each night where up
to $90$\% of coincidences between CA01 and CA05 were also coincident with a
CA03 event, indicating an RFI source.  The obvious
conclusion therefore is that the vast majority of observed
three-fold coincidences do not occur purely randomly, but rather
are triggered from a common event with significant
time-structure. By extension, there will be many such events seen
only in two antennas, and the same must therefore apply to the
two-fold coincidences, of which there are (generally) many more.

\begin{figure*}
\begin{center}
\includegraphics[width=0.85\textwidth]{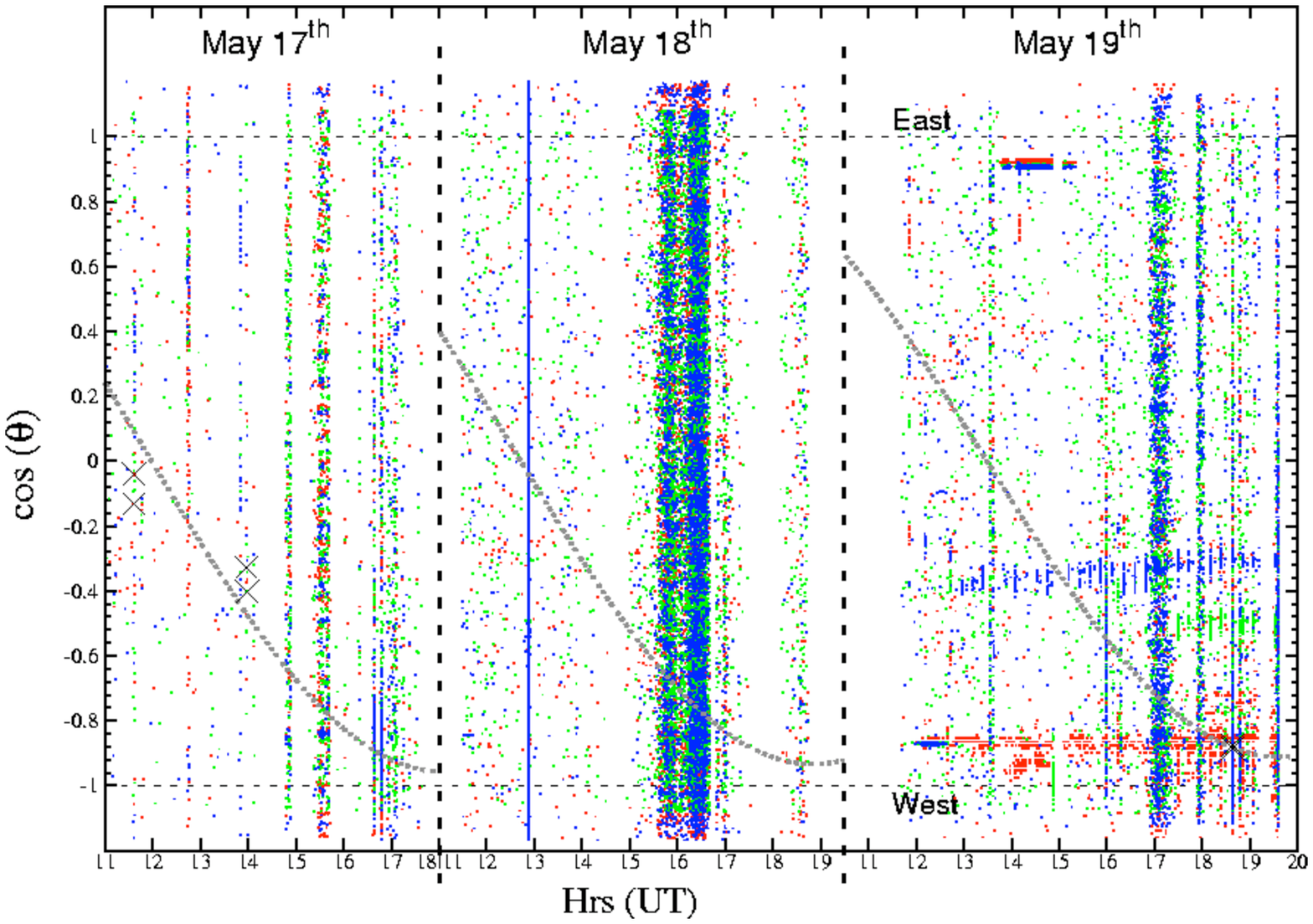}
\includegraphics[width=0.85\textwidth]{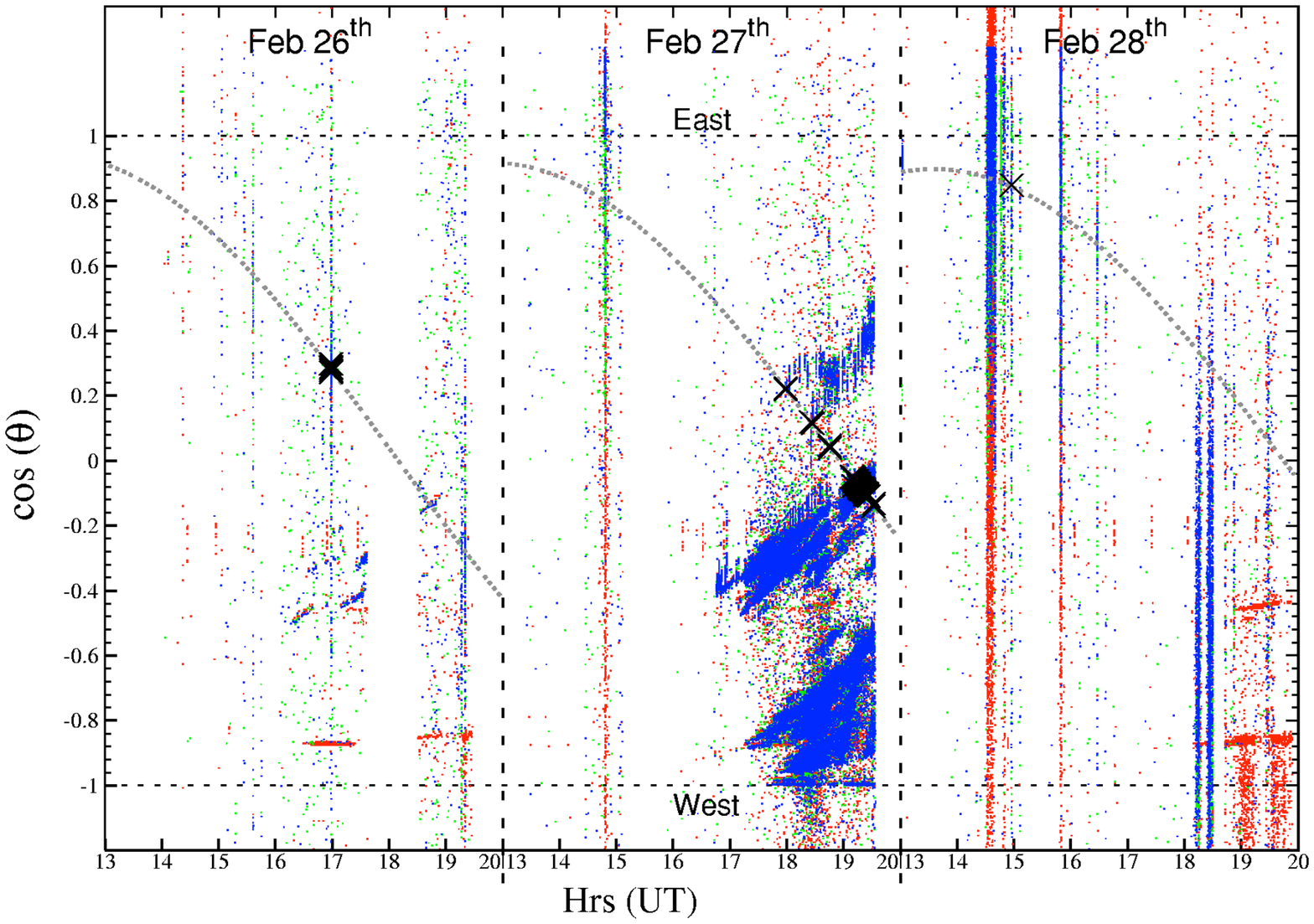}
\caption{
(color online) For each two-fold coincident event the angle,
$\theta$, between the East direction and the direction of
propagation of a plane wave fitting the arrival times at the two
relevant antennas is calculated, and $\cos\theta$ is plotted
(dots) against the time of the occurrence of each event for
February (top), and May (bottom). Red dots represent coincidences
between antennas CA01 and CA03, green dots for CA01 and CA05, and
blue dots for CA03 and CA05. In dot-dense regions, only blue
points show, since they are plotted last. The lunar direction (grey dotted line)
is also plotted, and positions of candidate events (crosses) are marked.
\label{ATCAtriple_ATCAevents}}
\end{center}
\end{figure*}

In Fig.\ \ref{ATCAtriple_ATCAevents}, we have plotted all
two-fold coincidences for each of the three pairs of antennas
over the entire observing period for both runs -- over $5 \times
10^5$ data points. For each two-fold coincident event the angle,
$\theta$, between the East direction and the direction of
propagation of a plane wave fitting the arrival times at the two
relevant antennas is calculated, and $\cos\theta$ is plotted
(dots) against the time of the occurrence of the event (top)
February, and (bottom) May 2008. Time increases continuously
except in breaks between days as indicated by the thick dashed
vertical lines.
Note that regions of high `dot-density' appear blue only because
the blue dots were plotted last. The lunar direction is indicated by
the grey dotted line, and positions
of candidate events (crosses) are marked.  For May 17th,
a network error necessitated many adjustments to the buffer size,
resulting in multiple unknown timing offsets to occur during that night.
Therefore the timing criteria for candidate lunar events have been
relaxed to $\pm 5$ sample offsets (i.e.\ $\pm$0.47 $\mu$s) on all antennas for May 17th,
explaining why the four candidate events for that night do not
lie exactly on the Moon's trajectory. For May 18 the trial offsets were
$\pm 1$ sample offsets (i.e.\ $\pm$94 nanoseconds) on all antennas, the system was more
stable and had two calibrations in agreement, and on May 19 (and all of February) the
times were completely aligned.  

A stationary source of pulse-like RFI producing triggers over a
long period of time will show up as a horizontal line, while a
very brief period of strong, narrow-band RFI will be observed as
a very large number of coincidences over a small time range but
large vertical extent (since the times will be random), i.e.\ a
vertical line.  

Fig.\ \ref{ATCAtriple_ATCAevents} provides an amazing amount of
information. The features can be approximately classed as below:

\begin{enumerate}
\item Short time periods exhibiting a high rate of coincident
triggers for all $\cos\theta$, mostly in May (e.g.\ 15:35 UT May
$17^{\rm th}$, 12:50 UT May $18^{\rm th}$, 10:00-17:20 UT May
$19^{\rm th}$). These appear as vertical features which extend
uniformly over the full range of offsets.
\item Short time periods exhibiting a high rate of coincident
triggers for a broad range of  $\cos\theta$, mostly in February
(e.g.\ 14:25 UT February $26^{\rm th}$, 14:40 and 14:50 UT
February $28^{\rm th}$, and 19:40 UT May $19^{\rm th}$). These
appear as vertical features which do {\it not} extend over the
full range of offsets.
\item Purely horizontal, typically thin features occurring at a
characteristic offset, sometimes over many days (e.g.\ the line
at $\cos\theta=-0.88$ for 17:00 UT on February 26th and
$\cos\theta=0.92$ for 14:30 UT on May 19th.
\item Sloped, typically broad features occurring most obviously
around 17:00--19:00~UT on February $27^{\rm th}$, but also on
February $26^{\rm th}$.
\end{enumerate}
Type $1$ features are exactly what would be expected from a high
random trigger rate, with triggers evenly spread in time-offsets.
The cause of the increased trigger rate must be a lowering of the
effective threshold by an increase in the background containing
no timing information, probably from narrow-band RFI -- an increase
due to ground temperature or the galactic background would be
unlikely to produce such short-duration bursts. As would be expected,
the triples rate is much lower than the doubles rate during these times,
due to the random nature of the trigger times.

Type $2$ features show both a small increase in triggers at all
offsets corresponding to a random component from an increased background,
and a large increase in triggers spread about a specific offset. Some
bands likely correspond to the nearby towns of Narrabri (East) and Wee
Waa (West-Northwest) which present potentially large, extended sources
of RFI. While these may be too weak to be detected under normal conditions,
a decrease in the effective threshold due to the presence of narrow-band
RFI could make such sources detectable. Another explanation is a single
source of RFI with a broad time-structure, with a high likelihood of
different antennas triggering off different parts of the signal, thereby
adding a random component with a preferred direction. These features
appear strongly for both doubles and triples, indicating the non-randomness
of the timing.

Type $3$ features act exactly as fixed sources of short-duration
RFI. The fixed geometrical delay results in a common time-offset
regardless of antenna pointing position (and hence
time). 

Type $4$ features remain largely unexplained. The locus of the
coincident triggers (lines with constant slope) are consistent
with some RFI source moving with constant speed in $\cos \theta$
($\theta$ the angle w.r.t.\ the baseline), which is strong enough
to give a high rate of triple-coincidences, and in the far-field,
since the delays per unit baseline match. Since multiple features
are seen at once, there must be either many such sources all
moving in unison, or many multiple reflections keeping the same
(and extremely large) angular offsets over a broad period of
time. Also, the apparent motion is at the sidereal rate, but in
the opposite direction. One suggestion is that a far-field RFI
source is being observed over multiple signal paths due to
tropospheric ducting which is associated with inversion layers.
Reflections off the antennas
themselves cannot explain the rate of change of delay being
proportional to the baseline length, nor is the antenna size of
$22$~m sufficient to produce more than a $\sim 70$~ns change in
delays. We can be certain however it is not an equipment fault
due to the presence of the aforementioned type $3$ feature during
this time period.

The apparent anti-sidereal motion of the features may result from
either the real motion of far-field sources of RFI, or a series of
reflections off an extended object, allowing each reflection point
to move smoothly with time.  In the former case, a possible
candidate is a set of satellites in a medium Earth orbit
(altitude $\sim 20,000$ km), which should move west-to-east across
the sky at approximately the correct rate.  This orbital altitude is
occupied primarily by navigational satellites, such as those of the
Global Positioning System (GPS), with `L1' and `L2' carrier frequencies of 1575.42
and 1227.6 MHz respectively.  The positions of the GPS satellites
over the period of the experiment were checked from public ephemeris
data \cite{GPS_ephemerides}, and found to exhibit the expected
anti-sidereal motion, but they did not match the positions of the features.
A more extensive search of all satellite positions might uncover a suitable candidate however.

\subsection{Source identification with three-fold triggers}

If any on-site RFI sources are found, they could be deactivated
and/or shielded in time for follow-up experiments, since in many
cases these events dominate our trigger rates and limit
sensitivity.  Given three antenna positions on an East-West
baseline, we can solve for the source position to within a
North-South ambiguity, since an event some distance North of the
baseline would produce exactly the same time structure if its
location were directly South of the baseline by the same
distance. We break the three-fold coincidences into two types of
events: near-field and far-field.

Using the timing offsets, a search for both far-field and
near-field events was performed for each block of data in each of
the February and May observation periods. The majority of
near-field solutions occur in the very near-field (within 1 km of
the antennas), and the rates of both near-field and far-field
events are highly variable. In many cases point-like sources of
RFI are seen, both in the near-field and far-field, and it makes
an interesting game trying to align the positions of possible RFI
sources with those detected.  Probable sources of RFI that we
identified in this way include the residence, the
control-building/lab, the solar observatory, the lodge,
and either or both of the Ionospheric Prediction Service center or visitors center, all of
which are on the ATNF site at Narrabri.

\section{Sensitivity Calibration}
\label{ATCAsensitivity_calibration}

To simulate the sensitivity of this experiment and place limits
on a flux of UHE neutrinos, trigger levels at each antenna in
terms of real quantities must be known. The usual specification
for an experiment such as this is the detection threshold $E_{\rm
thr}$ in V/m/MHz (a V/m threshold divided by the bandwidth) in a
given polarization just prior to being received by the
antenna. For instance, the GLUE threshold was a maximum field
strength per bandwidth of $E_{\rm thr} = 1.23~10^{-8}$~V/m/MHz,
calculated by taking the threshold of $6.46~10^{-9}$~V/m/MHz in
each circularly polarized data channel, and accounting for
vacuum--receiver transmission and the splitting of power between
polarizations \cite{Williams04}. To calculate our V/m/MHz
threshold, we first had to perform our own calibration of the
antenna gain as a function of frequency (bandpass calibration),
since the automated ATCA measurement of $T_{\rm sys}$ using the
injected noise source only applies to the standard signal path
and over a small frequency range. Hence, another method had to be
used, as described in the section below. Also discussed below are
the effects of ionospheric dispersion, which while approximately
corrected for, still reduced our sensitivity to some degree.

\subsection{The calibration function $k(\nu)$}
\label{ATCAdefs}

For a wide-band experiment such as this, the signal is expected
to change significantly in strength over the band. Note that
unless otherwise stated, $\Delta \nu$ refers to the total
bandwidth from $1.024$-$2.048$~GHz recorded, although the
sensitivity outside the range $1.2$-$1.8$~GHz will be minimal.
Our sensitivity will change as a function of frequency due to the
antenna, the receiver amplifier bandpass, and the de-dispersion
filter.  Given that our threshold is set in terms of our 8-bit
sampling, we need to be able to convert from a signal at the antenna resulting from coherent Cherenkov emission from a lunar cascade, 
$E(\nu)$ (V/m/MHz), to the value of the received buffer $b(t)$ at
the peak of the pulse ($t=t_{\rm peak}$). From here on, we simply
call the values of $b$ ``ADC digitization units'', or ``a.d.u.'',
and all frequencies are in units of MHz. The relationship between
$E(\nu)$ and $b(t_{\rm peak})$ involves an unknown (but
determinable) function $k(\nu)$, which is required to calculate
the peak signal height (arbitrarily, occurring at time $t=0$),
defined as below:
\begin{eqnarray}
b (t_{\rm peak}) ~{\rm (a.d.u.)} & = & \int_{\nu_{\rm min}}^{\nu_{\rm max}} k(\nu) E(\nu) d \nu \label{ATCAbknueqn}.
\end{eqnarray}
This gives the conversion between real field strength at
the antennas and the measured units in the CABB ADC boards.
For coherent pulses away from the peak, and for incoherent
signals at all times, the integral on the RHS of Eq.\
\ref{ATCAbknueqn} should include a phase factor $e^{2 \pi i \nu t}$,
i.e.\ it is a Fourier transform. Therefore $k(\nu)$ can be more
simply defined with respect to the Fourier transform $b(\nu)$ of
$b(t)$ as per Eq.\ \ref{ATCAbknueqn2}:
\begin{eqnarray}
b (\nu) & = & k(\nu) E(\nu) \label{ATCAbknueqn2}.
\end{eqnarray}
For simplicity, a more useful measure is $\bar{k}$, being the mean over the bandwidth $\Delta \nu$ between $\nu_{\rm min}$ and $\nu_{\rm max}$:
\begin{eqnarray}
\bar{k} & = & \frac{1}{\Delta \nu} \int_{\nu_{\rm min}}^{\nu_{\rm max}}  k (\nu) \, d \nu.
\end{eqnarray}
The sensitivity of the experiment $E_{\rm thr}$, defined in terms of a threshold electric field strength per unit bandwidth (for the simplest case of a flat-spectrum pulse), can then be calculated by knowing the trigger threshold $b_{\rm thr}$ using Eq.\ \ref{ATCAsimple_cal}:
\begin{eqnarray}
\label{ATCAsimple_cal}
E_{\rm thr}(\rm \nu) ~{\rm (V/m/MHz)} & \approx & \frac{b_{\rm thr}(t)}{\bar{k} \Delta \nu}
\end{eqnarray}
for $\Delta \nu$ in MHz. Therefore, in this section we calculate separately $k(\nu)$, $\bar{k}$, and hence $E_{\rm thr}$ for each data channel (antenna and polarization) over the entire observation period.

In order to calculate $k(\nu)$, a measurement of a known flux $F(\nu)$ (W/m$^2$/Hz) is required. For an incoherent signal (random phases), the relationship between $F(\nu)$ and the electric field over a given bandwidth is given by:
\begin{eqnarray}
\int_{\nu_{\rm min}}^{\nu_{\rm max}} F(\nu) d \nu & = & E_{\rm rms}^2/Z_0, \label{ATCAfnuermseqn}
\end{eqnarray}
where $Z_0=\mu_0c$ is the impedance of free space, and the RMS electric field is measured by the antenna system as:
\begin{eqnarray}
E_{\rm rms}^2 & = & \frac{1}{\Delta t} \int_{t_{\rm min}}^{t_{\rm max}} E(t)^2 dt \label{ATCA_erms}
\end{eqnarray}
Parseval's theorem for the $E(t) \leftrightarrow E(\nu)$ transform tells us that:
\begin{eqnarray}
\int_{t_{\rm min}}^{t_{\rm max}} E(t)^2 dt & = & \int_{\nu_{\rm min}}^{\nu_{\rm max}} E(\nu)^2 d \nu \label{ATCA_parseval}.
\end{eqnarray}
Substituting Eq.\ (\ref{ATCA_parseval}) into (\ref{ATCA_erms}), and thence into Eq.\ (\ref{ATCAfnuermseqn}), we arrive at Eq.\ (\ref{ATCAfnuermsnueqn}):
\begin{eqnarray}
\int_{\nu_{\rm min}}^{\nu_{\rm max}} F(\nu) d \nu & = & \frac{1}{Z_0 \, \Delta t} \int_{\nu_{\rm min}}^{\nu_{\rm max}} E(\nu)^2 d \nu \label{ATCAfnuermsnueqn}.
\end{eqnarray}
Since this relationship holds for an arbitrary bandwidth, the integration can be eliminated (e.g.\ let $\nu_{\rm max} \rightarrow \nu_{\rm min}$), giving:
\begin{eqnarray}
F(\nu) & = & \frac{1}{Z_0 \, \Delta t} E(\nu)^2 \label{ATCAfenueqn}.
\end{eqnarray}
Using Eqs.\ (\ref{ATCAbknueqn}) and (\ref{ATCAfenueqn}), the flux $F(\nu)$ as seen by the relevant data channel can then be related to the required calibration constant $k(\nu)$:
\begin{eqnarray}
F(\nu) & = & \frac{1}{Z_0 \, \Delta t} \left(\frac{b(\nu)}{k(\nu)}\right)^2 \label{ATCAfenueqn2}
\end{eqnarray}
which can be rearranged to give $k(\nu)$:
\begin{eqnarray}
k(\nu) & = & \frac{b(\nu)}{\sqrt{F(\nu) \, Z_0 \, \Delta t}} \label{ATCAkfromflux}
\end{eqnarray}
Eq.\ (\ref{ATCAkfromflux}) states that we can obtain $k(\nu)$ from a known flux $F(\nu)$ and the Fourier transform $b(\nu)$ of the corresponding sampled output.  Note that the output is actually sampled discretely, so $b(\nu)$ is obtained indirectly, through a discrete Fourier transform. Note also that for incoherent thermal emission, we do not need to track the phase change in each antenna over the bandwidth, i.e.\ we are interested only in the magnitude of $k(\nu)$.

\subsection{The Moon as a flux calibrator}

For our calibrator, we chose the Moon. The lunar temperature
$T_M$ is stable to within a few degrees over the lunar cycle at
approximately 225~K in the $1$--$2$~GHz range (see Ref.\
\cite{troitskij_tikhonova70}).  There are small errors in this
assumption due to the variation (1--2\%) with lunar cycle,
comparable variation across the band, and variations across the
disk of the moon, and small polarization effects.  Combining
these errors, this method should be accurate to within $5$\% or
better, which is acceptable.

Under these approximations, the lunar flux $F_M(\nu)$~(Jy) captured by the beam (it will be half this in any given polarization channel) is given by:
\begin{eqnarray}
F_M(\nu) & = & {2 \, k_b \, T_M \, \nu^2 \over c^2} \, \int_{\Omega_M} \, {\cal B} [\theta(\hat{\Omega},\hat{\Omega}_p),\nu] \, d \Omega \label{ATCAmoonflux}
\end{eqnarray}
where $k_b$ IS the Boltzmann constant,
 and ${\cal B}(\theta,\nu)$ is the beam-power pattern of ATCA
 telescope dishes \cite{techmemo} at frequency $\nu$ at angle
 $\theta$ to the telescope pointing direction $\hat{\Omega}_p$,
 (i.e.\ $\theta$ is the angle between directions $\hat{\Omega}$ and
 $\hat{\Omega}_p$), and $\Omega_M$ is the solid angle subtended by
 the Moon.  The beam-power pattern of ATCA deviates only slightly from
 the Airy pattern of a 22 m diameter aperture.  In the present
 observations, $\hat{\Omega}_p$ was either the direction towards
 the center of the moon or towards the lunar limb.

\subsection{Measurements}
\label{ATCAmeasurements}

To obtain $b(\nu)$ by observing lunar thermal emission we took an
unbiased sample of data pointing both on and off the Moon by setting
the trigger level to zero, i.e.\ maximally triggering. The
received flux $F_M(\nu)$ from the Moon can be detected by
subtracting the measured bandpass $b_{\rm off}(\nu)$ when
pointing away from the Moon from the bandpass $b_{\rm on}(\nu)$
when pointing at the Moon's center. The pointing-position for the
off-Moon data was a position at similar galactic latitude far
from any strong sources in the ATCA catalog. We set the buffer
lengths to maximum for this procedure, since then the product of
trigger rate and buffer length is largest, and also we obtain the
best spectral resolution. This was done once every time the
configuration was changed.

\begin{figure*}
\begin{center}
\includegraphics[width=0.8\textwidth]{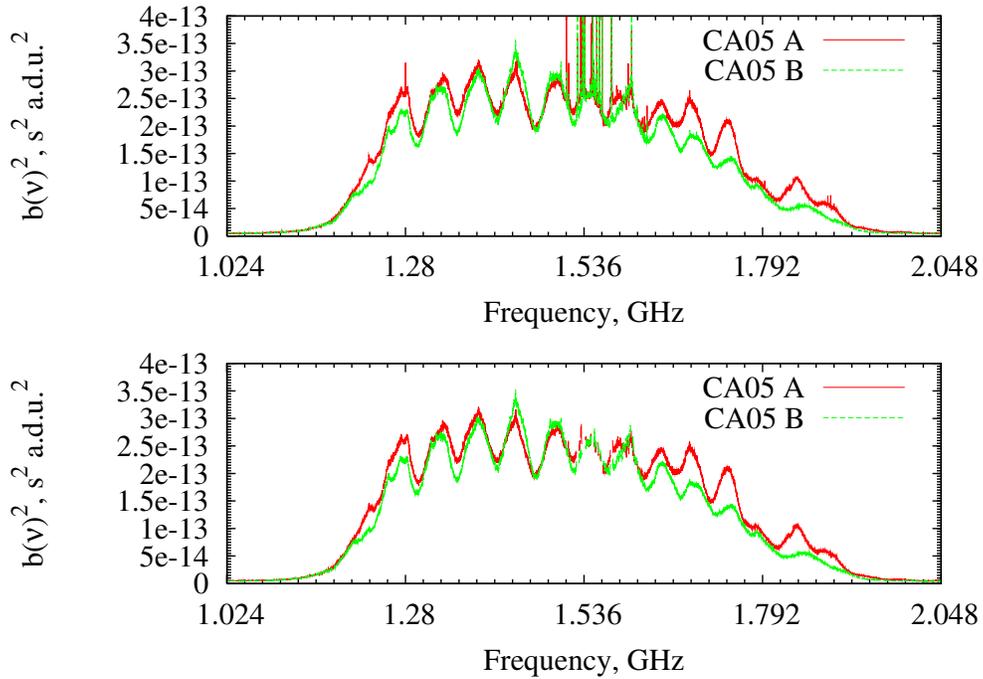}
\caption{(color online) Raw (top) and cleaned (bottom) spectra for the off-Moon pointing of May $18^{\rm th}$ on CA05. 
\label{ATCAraw_clean_spec}}
\end{center}
\end{figure*}

Each of the $N_b$ recorded buffers was
discrete-Fourier-transformed to produce $b(\nu)$ (typically $N_b
\sim 5000$). The resulting spectra are squared and then averaged
over all the buffers recorded for each calibration period/target
taken.  Each averaged spectrum is then cleaned with a very simple
cleaning algorithm to remove the worst of the RFI, which simply
sets the power of all RFI spikes above a running threshold to
zero, and the subsequent analysis ignores them. An example of the
raw and cleaned spectra is given in Fig.\
\ref{ATCAraw_clean_spec}.

The squaring, summing, and cleaning process was repeated for both
the off-Moon and on-Moon (center and limb) spectra for each
antenna/polarization, and the off-Moon power-spectrum is
subtracted from the corresponding on-moon spectra. Taking the
square root gave the required $|b(\nu)|$ corresponding
to the lunar contribution as required for Eq.\
(\ref{ATCAkfromflux}). Note that we make the approximation $|b(\nu)|\simeq b(\nu)$ assuming that the de-dispersion filter keeps the phase constant across the band.  This was then divided by the product
$\sqrt{Z_0 F_M(\nu) \Delta t}$ to give $|k(\nu)|$.

\subsection{Results of the calibration}

\subsubsection{Fitting for $k(\nu)$}
\label{ATCAknufitting}
\begin{figure*}
\begin{center}
\includegraphics[width=0.7\textwidth]{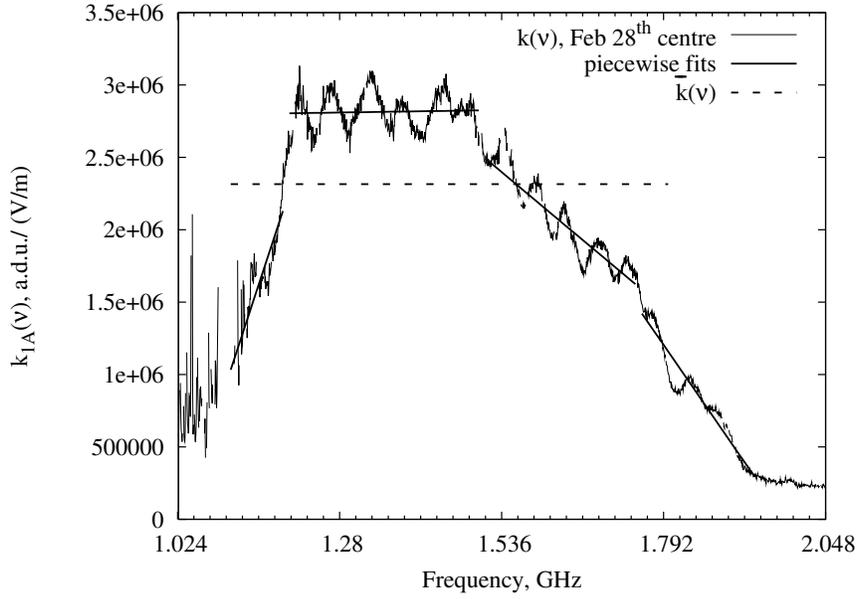}
\caption{$k(\nu)$ as measured for February 28$^{th}$ for CA01~A based on lunar center/off-Moon, piece-wise fits to $k(\nu)$.  Dashed line shows the mean $\bar{k} (\nu)$.
\label{ATCAca01_fitting}}
\end{center}
\end{figure*}

In order to characterize $k(\nu)$ in a meaningful way, a
piece-wise linear approximation to $k(\nu)$ was performed. Fig.\
\ref{ATCAca01_fitting} shows the fits for CA01~A in February 2008
-- four fits have been used, with different frequency ranges for
each of the February and May periods. Also shown is the mean
$\bar{k}(\nu)$, which has been fitted to the bandwidth
$1.1$-$1.8$~GHz in the case of February. For simulation
purposes, the piecewise-linear fit to the bandwidth was used.
Note that we have smoothed over the oscillations in $k(\nu)$,
which are caused by interference between the dispersion-corrected
signal from the filters, and a small reflection from the filter
connection point -- this reflection is also the cause of the pre-pulse
observed in Fig.\ \ref{ATCAnoise_cal_pulse}.

While some sensitivity
is not included by limiting the range of the fitted bandwidth,
including this range in the fit would artificially reduce
$\bar{k}(\nu)$ at lower frequencies where a signal is more likely
to be observed. Conversely, taking the fit below the
low-frequency cut-off would have led to an overestimate of the
sensitivity at low frequencies where the signal is
stronger. Unsurprisingly, the ranges which gave a good trade-off
between these effects and artificially reducing the effective
bandwidth $\Delta \nu$ were close to the nominal bandwidth of
$1.2$-$1.8$~GHz.

Only one calibration was performed in February, and no cross-checking was possible. In May, it was found that while fits from both May $19^{\rm th}$ measures and the limb-off-Moon fit of May $18^{\rm th}$ were always in good agreement ($\pm 3$\%), the lunar center/off-Moon fits of May $18^{\rm th}$ were consistently low by $5$-$8$\%. Therefore these were excluded, and the fits averaged over the remaining data. For February, it may therefore be that a similarly low (or high) result was obtained. Such a systematic error is nonetheless small compared to other uncertainties in our sensitivity calculation, so we do not carry this error through our calculation.

\subsubsection{Conversion to an effective $V/m/MHz$ threshold}
\label{ATCA_sensitivity_real_units}

Using the piecewise linear approximations to $\bar{k}(\nu)$ it is possible to calculate the thresholds in V/m/MHz given the thresholds in a.d.u. These were constantly altered throughout the experiment. The thresholds varied significantly between data channels and over time, since the thresholds were chosen to keep the trigger rate on each channel (rather than the thresholds themselves) constant.

Since a positive detection requires a three-fold trigger, it is
useful to define an effective signal detection threshold over all antennas. A good
measure is to choose a mean V/m/MHz signal strength that over the
bandwidth will have a 50\% probability of triggering all three antennas.
The random noise component can either increase or
decrease the measured signal, so we require sufficient intrinsic
signal strength above each individual antenna threshold that the chance of the random
component pushing the signal below threshold is small. Thus the
effective threshold is dependent upon this random
component. The individual antenna thresholds were calculated from the mean
recorded spectra from the relevant limb/center pointing
calibrations by averaging the RMS signal over the entire
$1.024$~GHz bandwidth, and converting the measured RMS signal in
a.d.u.\ into V/m/MHz using $\bar{k}$. During times of significant
out-of-band RFI, the effective thresholds will vary due to a
greater RMS signal, but since these occasions are both rare and
have a low effective efficiency, their contribution to the
average threshold will be negligible and is neglected here. 

Assuming a normally-distributed RMS field strength, the
probability of the total of signal plus noise falling above
threshold for any given signal strength can be readily
calculated, and thus the probability that the global condition
{\small \tt (CA01A OR CA01B) AND (CA03A OR CA03B) AND (CA05A OR
CA05B)} will be met. The detection probability is thus dependent on
the alignment of the field vector with the A and B receivers --
since the trigger condition is {\small \tt A OR B}, the
probability is highest when the field vector is parallel with
either the A or B polarization directions, and lowest when it is
$45^{\circ}$ from both.  Therefore the `effective threshold' is
defined for a signal polarized at $22.5^{\circ}$ (i.e.\ half-way
between $0^{\circ}$ and $45^{\circ}$) to either A or B. Since in
general both the thresholds and RMS values are different for each
polarization, we calculate the effective thresholds for signals
polarized at both $22.5^{\circ}$ ($67.5^{\circ}$) and
$67.5^{\circ}$ ($22.5^{\circ}$) to the A (B) receiver directions
and average the results. By varying the raw signal strength
until the calculated three-fold detection probability was $50$\%,
the effective thresholds could be calculated.

Doing so, we found that whereas each individual
antenna trigger threshold was in the range $1.1$-$1.3$ $\times 10^{-8}$
V/m/MHz, the effective thresholds for a global trigger were
in the range $1.45$-$1.6$ $\times 10^{-8}$ V/m/MHz, i.e.\ an
increase in threshold (decrease in sensitivity) of approximately $25$\%.
In comparison, adding all three antenna voltages coherently (gain of $\sqrt{3}$
in signal-to-noise) and detecting at a higher level of $V_{thresh} = 9.5 V_{\rm RMS}$
(so the probability of a false detection for a $\sim30$-hr experiment
would be less than $0.1$\%; there would no longer be a coincidence check)
would have produced an effective threshold a few percentage {\it less}
than the individual antenna thresholds, i.e.\ $1.05-1.25$ V/m/MHz.
Thus our inability to combine the signals coherently reduced our sensitivity
by approximately $30$\%.

\subsubsection{Comparison with a possible experiment at Parkes}

An alternative instrument to the ATCA is the Parkes $64$~m single dish radio
telescope, with an effective bandwidth of $300$~MHz in the $1.2$-$1.5$~GHz range.
While the total sensitivity (here, area-bandwidth product) compared to the six
ATCA antennas at $600$~MHz bandwidth is $30$\% lower, with current
technology we are unable to take advantage of the full ATCA collecting area,
so we present a brief comparison of the two instruments.  Approximate
scaling from ATCA to Parkes would suggest that: (i) we gain a
factor of $\sim$9 in area compared with one $22$~m dish; (ii) we
loose a factor of $\sim$2 because we don't have a triple
coincidence trigger and so must set the trigger threshold higher;
(iii) we loose a factor of $\sim$2 because of the smaller
bandwidth at Parkes; (iv) there is a modest gain because the
lowest frequency is 1~GHz and not 1.2~GHz; (v) there is a modest
gain because of the higher fraction of high quality RFI-free
on-Moon time, though a lack of baseline may make RFI discrimination
more difficult; (vi) there is a loss of a factor $2$--$3$ because the Parkes
multibeam receiver can cover less of the lunar limb. Factors (i)--(iv)
reduce the neutrino energy threshold by $\sim \sqrt{9/(2\times 2)}$,
i.e.\ overall the sensitivity for our ongoing experiment at Parkes should
be more than twice as sensitive as our 2008 experiment using
the ATCA, while factors (v)--(vi) decrease the effective area
to high-energy events by $50$\%. For targeted observations of potential
sources of UHE particles (see Ref.\ \cite{JamesProtheroe_DirectionalAperture2009}),
the Parkes telescope will be even more suitable, since for a targeted observation,
only part of the lunar limb would need to be observed.


\section{Results}
\label{ATCA_results}

\subsection{Search for lunar pulses}

The main search criteria used for eliminating false events were
the timing requirements. The search window is given by the
apparent angular width of the Moon in the East-West direction,
since the North-South component is unresolved by the East-West
baseline. This is $\sim 0.5^{\circ}$ at transit (when the Moon
achieves its greatest elevation), and considerably less near
Moon-rise/set. This gives an intrinsic time-window of up to
$23$~ns ($46$ samples) over the maximum baseline of
$765$~m. While neither the raw nor the correlation-corrected
times can be completely trusted, true lunar pulses will have a
sharp time structure which will allow only a small variation in
trigger times between antennas. Alternatively, those
with extended structure (i.e.\ multi-peaked electromagnetic
showers viewed away from the Cherenkov angle) will be due to the
highest energy showers and therefore strong enough to give a
correct correlation.  

Performing the search over both observation periods resulted in
60 candidates.  Note that for any given lunar position there will
always be two points on the horizon (one North and one South)
which will have the same timing solution for an East-West
baseline.  Our search criterion is illustrated graphically in
Fig.\ \ref{ATCAtriple_ATCAevents} (see Sec.\ \ref{RFI}), where
both the apparent direction of the Moon and the times and origins
of candidate signals are plotted. Candidate origins plotted
for May 17 do not appear to be consistent with the lunar position
because we have allowed a larger time window in the search due to
an uncertainty in the timing calibration for that night (as previously
noted). Note that all candidate events were detected during
periods of intense RFI.

The candidate events were then searched through by eye for
pulse-like events, and it was found that a majority of
the candidates had a narrow-band RFI signature, with the recorded
time-domain signals being strong over the entire buffer length.
We did not use a more quantifiable measure than `pulse-like' simply
because the narrow-band RFI was so obvious and strong when present, and
some `pulses' had duration up to $30$~ns. Since at these extremely high energies,
multiple cascade signatures (e.g.\ from nearby hadronic and purely electromagnetic cascades)
might cause a lengthening of the expected pulse profile, we preferred to use
timing criteria only if possible.

\begin{figure*}
\begin{center}
\includegraphics[width=\textwidth]{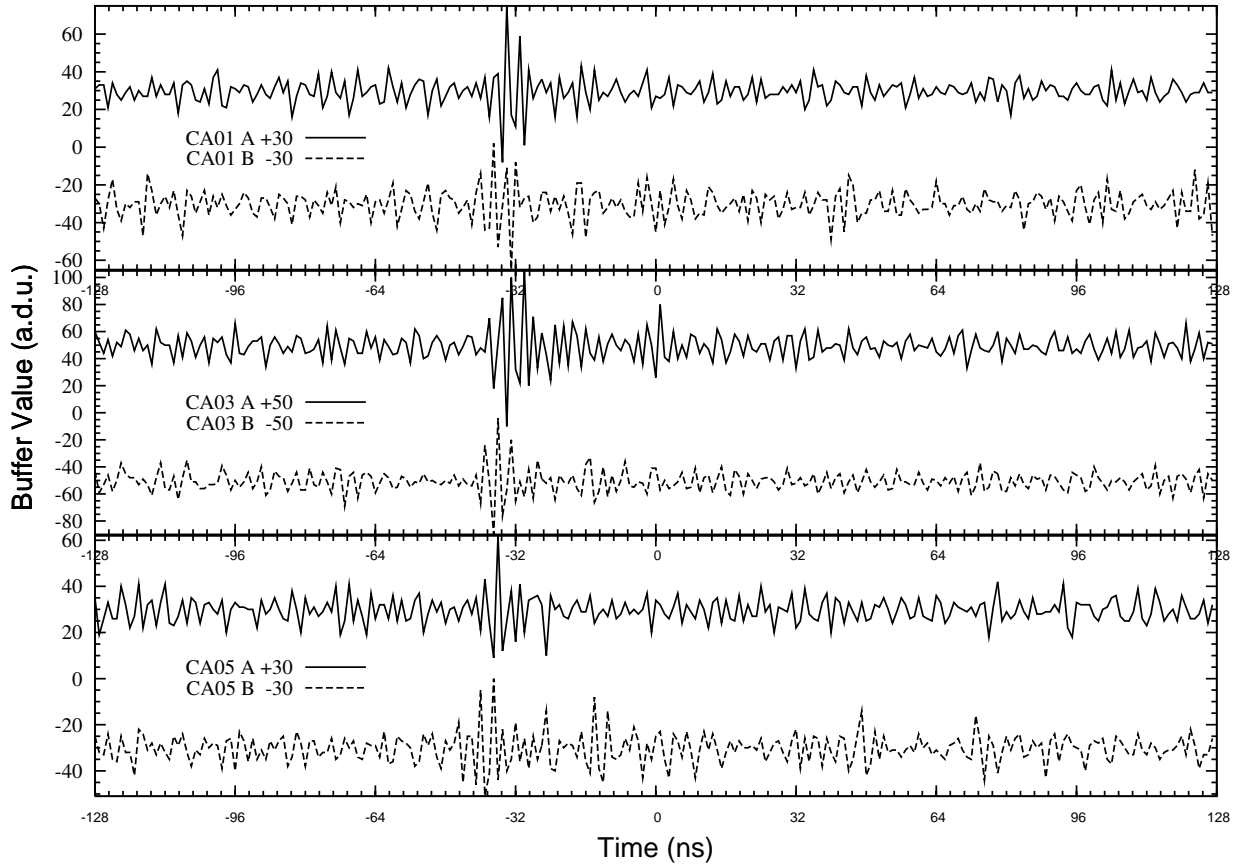}
\caption[A narrow time structure event from February $27^{\rm th}$.]{A narrow time structure event from February $27^{\rm th}$.  Note that the signal has been displaced vertically by $\pm$30 as indicated for display purposes.}
\label{ATCAlikely_threefold}
\end{center}
\end{figure*}

A minority of events
-- $16$ in total -- had a narrow time-structure, an example of
which is given in Fig.\ \ref{ATCAlikely_threefold}. All came
within a two-hour period on February $27^{\rm th}$, which was one
of the most RFI-intense periods of all the observations. These
could not be immediately excluded by eye, and had to pass more
stringent tests. These are described below.

\subsubsection{Ensuring possible origin to within sampling accuracy.}
\label{ATCAfarfield_byeye}

\begin{figure*}
\begin{center}
\includegraphics[width=0.7\textwidth]{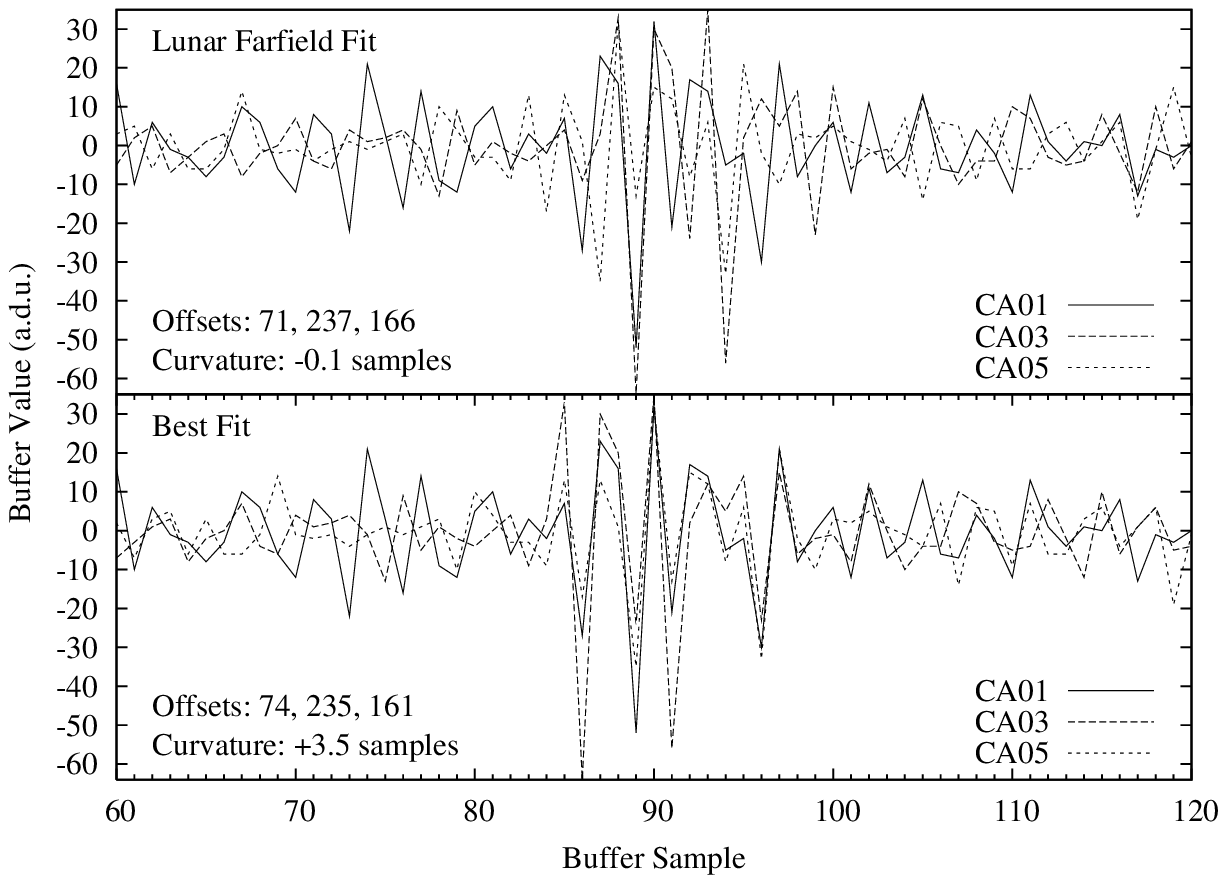}
\caption[Best by-eye alignment of a candidate event.]{Comparison of best far field-fitted alignment (top) with the best unrestricted fit (bottom) for polarization B on one of the sixteen narrow-time-structure candidates on February $27^{\rm th}$, for the values of the offsets ($\Delta t_{31}$, $\Delta t_{51}$, $\Delta t_{53}$) shown. The `wavefront curvature' is given by subtracting the measured $t_3$ from the expected value $t_3^{\prime}$ given by Eq.\ \ref{ATCAt3expected}. }
\label{ATCA_farfield_likely}
\end{center}
\end{figure*}

The search algorithm allowed both small deviations from a
far-field event, and small offsets in direction from the Moon, to
account for potential errors in the automated alignment
process. A check by eye of the corrected alignment times, if
necessary including further adjustments, would be expected to
yield accurate timing information in cases where the detected
event has significant time-structure, as is the case with all
$16$ candidates. The correlation-corrected times occasionally
needed further adjustment by one sample on a single baseline. For
each candidate, the resulting alignment was compared visually to
that required for the event to have a far-field origin; this was
done quantitatively by comparing the buffer trigger times for
{\sc CA03}, $t_3$, to that expected ($t_{3}^{\prime}$) from $t_5$
and $t_1$:
\begin{eqnarray}
t^{\prime}_3 & = & \left( (t_5-t_1) (b_{13}/b_{15}) + t_1 \right) \label{ATCAt3expected}.
\end{eqnarray}
where $b_{13}$ and $b_{15}$ are baselines between antennas CA01
and CA03, and CA01 and CA05, respectively.  An example of this
procedure is given in Fig.\ \ref{ATCA_farfield_likely}. In no
case did any of these events appear to be a far-field event. In
most cases, structure was evident in both polarizations, but was
so weak in one that the other only could be used for determining
the alignment -- however, an alignment to within sampling
accuracy could always be obtained. Since at these times the Moon
made an angle of nearly $90^{\circ}$ to the ATCA baseline, events
within $360$~km would result in a wavefront curvature with
measurable differences between $t_3$ and $t_3^{\prime}$ -- in the
case shown in Fig.\ \ref{ATCA_farfield_likely}, the distance is
of order $100$~km and so this event can not have a lunar origin.

In principle we could have also used the expected dispersion
measure as a test to verify a pulse being of lunar or terrestrial
origin.  However, we deliberately chose observing times when the
dispersion was low, and so this reduced the power of such a test.
Since our timing criteria were already sufficient to exclude all
candidate events as RFI we did not pursue this approach.

\subsection{Effective apertures to an isotropic flux}

The simulation program described by James \& Protheroe
\cite{JamesProtheroeLCT109} was modified to weight the frequency
spectrum output over the bandwidth by $k(\nu)/\bar{k}$ using the
piece-wise linear approximation.  The sensitivity of the
experiment did not remain constant, but this was taken into
account by running simulations using the lowest and highest
values of $E_{\rm thresh}$ for each of the observation periods.

The deviation of the true TEC compared to that designed into our
de-dispersion filters changed our sensitivity continuously. The
average peak recorded signal strength as a fraction of intrinsic
peak strength (Fig.\ \ref{ATCAall_tecu_graph}) for each of the
observation periods was folded into the simulations for different
combinations of the two extreme Cherenkov spectra and the lowest,
average, and highest sensitivity.

It was found that changing the angles of the polarized receivers
with respect to the lunar limb varied the calculated isotropic
apertures to neutrinos above $10^{21}$~eV by less than $1$\%,
and at most $4$\% at $10^{20}$~eV, where the apertures are in
any case very small.

The two major uncertainties in this calculation are the UHE
neutrino interaction cross-sections, and the effects
of small-scale surface roughness (SSR), both of which are dealt with
in detail in the appendices. In the following results, we do not incorporate
the uncertainty in the cross-section for the simple reason
that the uncertainty itself is so uncertain, since the scope
for new physics at such high center-of-mass-energy collisions
is large. Rather, we summarise the results of Appendix \ref{xsectionapp}
by stating that a doubling (halving) of the cross-section
results in an increase (decrease) in the effective aperture 
by a factor of approximately $1.88$.

For the effects of small-scale surface roughness, 
we develop a toy model, and show both standard estimates (``ATCA no SSR'') and
estimates adjusted for the results of calculations using this model
(``ATCA SSR''). This model, while useful for understanding
SSR phenomenology, is much less sophisticated than this topic requires,
and the results based on it should be
interpreted more as guides to indicate the nature of small-scale
surface roughness effects. The important result is that whereas
traditionally the effects of SSR were thought of as entirely negative,
since the effect is to reduce the coherence of the wavefront over the
cascade length, the resulting increase in angular spreading of the radiation
makes a detection at the highest energies more likely. Therefore, the calculations
including SSR have a lower effective aperture at low neutrino energies, and
a much higher aperture at the highest energies. Future work is expected
to provide a more quantitative estimate of SSR effects.

\subsubsection{Apertures and limits to an isotropic flux}

\begin{figure}
\begin{center}
\includegraphics[width=0.5\textwidth]{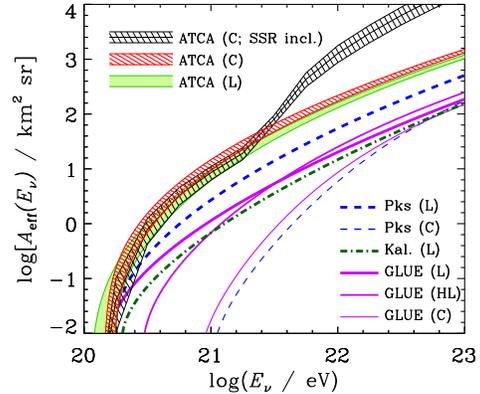}
\caption{(color online) The range of effective apertures (see
text) for the LUNASKA February 2008 center-pointing (C) and May
2008 limb-pointing (L) ATCA observations, compared to that from
previous experiments at Parkes (`Pks'), Kalyazin (`Kal.'), and
Goldstone (`GLUE'), in both limb- (`L'), half-limb- (`HL'), and center- (`C')
pointing modes (see \cite{JamesProtheroeLCT109}), assuming
the existence of a sub-regolith layer of comparable dielectric
properties to the regolith itself. We also show the effect on the
aperture for the (center-pointing) Feb.\ 2008 observations
including our adjusted toy model of small-scale surface roughness
(C, SSR incl.) (see Appendix \ref{roughness}), which is the reason
for the abrupt (and artificial) change in aperture near $2 \times 10^{21}$~eV.}
\label{ATCA_isotropic_app}
\end{center}
\end{figure}

The resulting range of effective apertures to an isotropic flux
from each period is given in Fig.\ \ref{ATCA_isotropic_app},
assuming the presence of a radio-transparent mega-regolith.
Also plotted are the effective apertures
from prior experiments as calculated by James \& Protheroe
\cite{JamesProtheroeLCT109}. For our ATCA observations, the
threshold is lower than in past lunar Cherenkov experiments
(since our high bandwidth has compensated for the smaller dishes),
while the effective aperture at high energies is greater (due to
increased coverage of the lunar limb and lower
frequencies). Unlike the previous experiments with larger dishes, the
sensitivity to UHE neutrinos is higher in the center-pointing
mode (Feb $2008$) than in the limb-pointing mode (May $2008$), since
the beam-width of ATCA near $1$--$2$~GHz is comparable to the
apparent diameter of the Moon. Note also that previous
calculations (Refs.\ \cite{Williams04, Beresnyak04, Scholten06})
have not included the loss from a non-infinite sampling rate, and
so their effective aperture should be reduced somewhat.  

\begin{figure*}
\begin{center}
\includegraphics[width=0.7\textwidth]{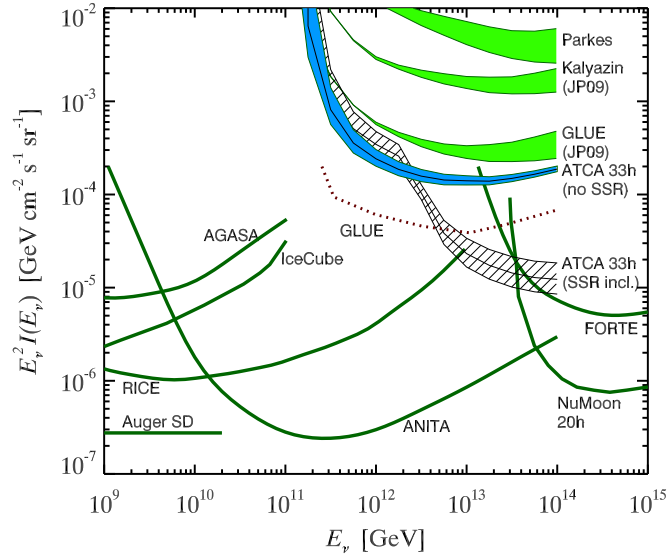}
\caption[Limit on an isotropic flux of neutrinos from LUNASKA
ATCA observations.]{(color online) Model-independent $90$\% confidence limits, i.e.\ $2.3 / [t_{\rm
eff} A_{\rm eff}(E_\nu)]$ for our nominal $33.5$~hr observations
(effective on-time $t_{\rm eff}=26.15$~hr), on a total flux of UHE neutrinos
(adjusted for all neutrino flavours) from our 2008 ATCA
observations assuming a sub-regolith layer, both using the 
standard modeling (``ATCA no SSR'') and using our adjusted toy model of 
small-scale surface roughness (``ATCA SSR incl.''). In the `SSR' case,
the abrupt transition near $2 \times 10^{12}$~GeV is a model artefact
(see Appendix \ref{roughness}). Also, from previous experiments: GLUE \cite{Gorham04}; IceCube
\cite{IceCube07}; RICE \cite{Kravchenko06}; ANITA 
\cite{GorhamANITA08}; FORTE \cite{Lehtinen04}; NuMoon \cite{NuMoon20hARENA2008}; revised estimates
by James \& Protheroe \cite{JamesProtheroeLCT109} for Parkes, GLUE, and Kalyazin are
shown by hatched bands (upper boundary -- limit for 10~m
regolith; lower boundary -- 10~m regolith plus 2~km
sub-regolith); Auger surface detectors \cite{auger_nu}. The range
on the band labeled ATCA reflects experimental uncertainties,
while for previous experiments (where applicable) reflects the
inclusion or otherwise of a sub-regolith layer. }
\label{ATCAisotropic_lim}
\end{center}
\end{figure*}

The limit on an isotropic flux of neutrinos arising from our
combined 2008 observations to a UHE $\nu$ flux is given in Fig.\
\ref{ATCAisotropic_lim} as the band labeled ``ATCA no SSR''.
Also shown is the limit -- ``ATCA SSR'' -- when our toy model of small-scale surface roughness
(see Appendix \ref{roughness}) is included, and the
limits from previous experiments, including GLUE
\cite{Gorham04} (dashed line labeled ``GLUE'') which is now
believed to be approximately an order of magnitude too low as
pointed out by James \& Protheroe \cite{JamesProtheroeLCT109} and
confirmed by Gayley et al.\ \cite{GayleyMutelJaeger2009}.  The
GLUE limit as revised upward  by James \& Protheroe is shown by the band
line labeled ``GLUE (JP09)''.

\section{Summary and Conclusion}
\label{ATCA_conclusion}

In 2008 we carried out observations of the Moon at the Australia
 Telescope Compact Array using the lunar Cherenkov technique to
 search for the signatures of UHE neutrinos. Although no lunar
 pulses of any source were positively identified, the
 instantaneous apertures for all the observations were the most
 sensitive yet achieved using the technique. The corresponding
 limits on an isotropic flux are not as strong as those from
 RICE and ANITA, but our methods to
 improve sensitivity to certain patches of the sky --
 which, along with our limit on the UHE neutrino
 flux from Centaurus A and the Galactic Center
 \cite{ATCA2008data_CenA_GC_limit}, will be published
 separately -- were a
 success: we found the exposure from our observations to Centaurus
 A to be higher than all previous experiments at neutrino energies
 of $10^{22}$--$10^{23}$~eV and above.

The lack of detection of nanosecond lunar pulses reported here is
consistent with limits set by the ANITA experiment. Importantly,
our methods to discriminate between true lunar pulses and RFI
were very successful: we can be extremely confidant 
that no true lunar Cherenkov pulses were detected simultaneously by all three
antennas, despite having of order ten million trigger events. This demonstrates primarily
the power of nanosecond timing over a significant baseline, which
was made possible by our use of a very wide (here, $600$~MHz) bandwidth,
and was the main criterion used to discriminate against false
events. Importantly, it was found in Sec.\
\ref{ATCAfarfield_byeye} that for events of reasonable
time-structure (the category into which all candidates must
fall), the automatic procedures produced times accurate to better
than $1$~ns. Therefore, relying on such a procedure to search for
candidate pulses in the future is justified, making a comparable
analysis for a long observation run of a month or
more feasible.

Our first estimates of the effects of small-scale surface roughness
on the detectability of lunar Cherenkov pulses at high frequencies
--- and the potential two orders of magnitude improvement in the effective
aperture at the highest energies, and order-of-magnitude worsening of
the threshold --- only serves to emphasize the importance of further work.
Since these effects are indicative of `messy' signals caused by
roughness-induced interference along the length of the cascade,
it also suggests that timing criteria on signal origin (i.e.\ that
the signal comes from the Moon) be used instead of criteria on the
pulse shape (i.e.\ that it is very close to impulsive). Our use
of such criteria to eliminate all our candidate events proves
that this is possible. It is important to note that the energy range
at which small-scale surface-roughness effects are expected to increase
the strength of our limit is the range at which our ATCA experiment
is most competitive.

For our observations, the use of an analog de-dispersion filter
proved highly successful. The dispersion measure assumed in the
filters' construction turned out to be very close to the actual
values during observation periods, so that incorrect
de-dispersion lost us less sensitivity than triggering
inefficiency due to our non-infinite sampling rate. Though such filters
must inevitably be superseded by a digital method, their
continued use in the meantime will be valuable. Conversely, the
finding that the loss from a finite sampling rate was greater for
our simple trigger algorithm than that from incorrect
de-dispersion is extremely important, and that in fact our
`over'-sampling was an important factor in increasing (or rather,
reducing the loss of) sensitivity. In future observations
therefore the use of a smarter trigger algorithm that uses the
already fully available information to reconstruct intermediate
trigger values should be used, and perhaps should take as high a
priority as digital de-dispersion.

Compared with an alternative single-dish experiment at Parkes,
our experiment at the ATCA provided more effective area to high
neutrino energies at the expense of less sensitivity to lower energy events.
Since the parameter space at which the ATCA experiment is superior
is best explored by low-frequency experiments such as NuMoon,
we have transferred our efforts to utilising the Parkes dish,
the results of which will be reported in a future contribution.
We point out, however, that multi-telescope systems
(such as the SKA and its pathfinders) will be more
sensitive than single-dish experiments, and that our ATCA
experiments would have been significantly more sensitive had we
not been limited to using only three antennas due to telescope
upgrade delays.

For future experiments at the ATCA or at other radio telescope
arrays, further improvements such as real-time coincidence logic
between three or more antennas, or even the ultimate goal of a
coherent addition of the signals, would also improve the
sensitivity. Without a further analysis of the typical RFI
structure, it is not possible to determine the utility of
real-time anti-RFI logic, though given the prevalence of
RFI-triggered pulses, this too should be considered.

The lessons learned above should in all cases be applicable to
any use of the lunar Cherenkov technique with an array of radio
antennas. The advantage of using a giant radio array such as the
SKA to search for lunar pulses has only been highlighted by these
observations, especially since it will be placed in a low-RFI
environment.

We have demonstrated techniques being developed by us ultimately
for use with the Square Kilometer Array, and have been able with
only 6 nights of observations using the ATCA to produce the
lowest limit to an isotropic UHE neutrino flux below $3 \times 10^{22}$~eV of any lunar
Cherenkov experiment.  While at present the isotropic limits from
lunar Cherenkov experiments are not competitive with RICE
\cite{Kravchenko06}, ANITA \cite{GorhamANITA08} and (above
$3\times 10^{22}$~eV) NuMoon, use of the SKA in several years'
time for lunar Cherenkov observations will provide a very
powerful technique for UHE neutrino astronomy \cite{JamesProtheroeLCT109}.
With an estimated sensitivity to neutrinos $100$ times less energetic
than those detectable with our experiment at the ATCA, the SKA will be able to
probe the as-yet untested predictions for a flux
of neutrinos from the GZK process.
Furthermore, our
current experiment has been able to access regions of the sky not
accessible to ANITA and NuMoon, and with better exposure than
RICE above $3\times 10^{22}$~eV. Our flux limits to UHE
neutrinos from Centaurus A will be discussed in a future
paper.

\appendix

\section{Variation of aperture with neutrino cross-section}
\label{xsectionapp}

The greatest unknown in the calculation of the aperture is that
of the UHE neutrino-nucleon cross-section $\sigma_{\nu N}$, since
this requires the extrapolation of experimental data over many
orders of magnitude -- hence the finding of Gandhi {\it et al.}
\cite{Gandhi98} that the cross-section could vary by a factor of
$2^{\pm 1}$ at $10^{20}$~eV \cite{Gandhi98}. This conclusion was
born out after a more recent calculation by Cooper-Sarkar and
Sarkar \cite{sarkar08} using updated data from HERA estimated a
neutrino-nucleon charged-current (CC) cross-section which is
approximately $30$\% lower at $10^{20}$~eV, and getting
relatively smaller with increasing energy. There is also scope
for `new physics' to alter the cross-section even further.

Given the range of uncertainty even within `standard physics',
the determination of the UHE neutrino-nucleon cross-section is
necessarily a scientific goal of UHE neutrino-detection
experiments which is inseparable from the measurement of the UHE
neutrino flux itself, and in this context, instead of limits on
the flux having a cross-sectional uncertainty, the limits should
be set in flux--cross-section space. However, due both to
convention and the complexity of doing so, they are not. Instead,
to estimate the effect of uncertainty in the cross-section on our
calculated effective area and flux limits, we calculate the
fractional rate of change in the effective aperture $A_{\rm eff}$
with the fractional rate of change in cross-section, i.e.\
$\frac{\sigma_{\nu N}}{A_{\rm eff}} \frac{d A_{\rm eff}}{d
\sigma_{\nu N}}$. In the limit where neutrinos cannot penetrate a
large part of the Moon and are seen only when they interact
almost immediately in a thin layer at the Moon's surface
(`down-going' interactions), we expect $\frac{\sigma_{\nu N}}{A_{\rm
eff}} \frac{d A_{\rm eff}}{d \sigma_{\nu N}}=1$, i.e.\ doubling the
cross-section doubles the interaction rate, and vice-versa (a
similar effect is reached if the entire Moon is transparent to
neutrinos, but this is far from reality). Since this limit is
approached for high neutrino energies observed at high
frequencies, we expect results to be close to $1$. Any
contribution from `up-coming' neutrinos (neutrinos having
penetrated a significant part of the Moon before reaching the
surface) would reduce the result below $1$, while no mechanism
exists to increase the result above $1$.

To determine the sensitivity to the cross-section, we varied the cross-section by $\pm 20$\%, and ran simulations for three primary neutrino energies -- $10^{21}$, $10^{22}$, and $10^{23}$~eV -- using the full range of ATCA sensitivities (best- and worst-cases) and configurations (February and May observations). We found the effect of varying the cross-section at a given energy was the same for all cases, and that over this range of $\sigma$, $\frac{\sigma_{\nu N}}{A_{\rm eff}} \frac{d A_{\rm eff}}{d \sigma_{\nu N}}$ was also constant at a given energy. Thus in Table \ref{xsectiontbl} we give one value only of $\frac{\sigma_{\nu N}}{A_{\rm eff}} \frac{d A_{\rm eff}}{d \sigma_{\nu N}}$ for each primary neutrino energy, averaged over all cross-sections and observer configurations. That the values are very close to one confirms that down-going neutrinos dominate the detected interactions, especially at high energies, so that a reduced (increased) estimate for the neutrino-nucleon cross-section would proportionately decrease (increase) the ATCA experimental aperture, with a corresponding increase (decrease) in the flux limit set from this experiment. Whether the cross-section can be deconvolved from the flux using (for instance) the average origin of the signal is a subject for a future contribution. We also give in Table \ref{xsectiontbl} the reduced values of the charged-current neutrino-nucleon cross-section $\sigma_{\nu N CC}$ using the fit given in Eq.\ 3.5 of Cooper-Sarkar and Sarkar (CSS) and Gandhi {\it et al.} (GQRS), together with the implied reduction in effective area $A_{\rm eff}^{CSS} / A_{\rm eff}^{GQRS}$ under the assumption that the neutral-current cross-section scales with the charged-current cross-section in the CSS calculation.

\begin{table}
\begin{center}
\caption{The effects of changing cross-section on the simulated effective aperture, assuming the neutral-current
cross-section scales with the charged-current cross-section for the calculation of Cooper-Sarkar \& Sarkar \cite{sarkar08}. \label{xsectiontbl}}
\renewcommand{\arraystretch}{1.9}
\begin{tabular}{lccc}
\hline
\hline
$E_{nu}$	& $10^{21}$~eV & $10^{22}$~eV & $10^{23}$~eV \\
\hline
$\frac{\sigma_{\nu N}}{A_{\rm eff}} \frac{d A_{\rm eff}}{d \sigma_{\nu N}}$ & $0.9$ & $0.915$ & $0.93$ \\
$\sigma_{\nu N CC}^{CSS} / \sigma_{\nu N CC}^{GQRS}$ & 0.55 & 0.42 & 0.30 \\
$A_{\rm eff}^{CSS} / A_{\rm eff}^{GQRS}$ & 0.60 & 0.46 & 0.35 \\
\hline
\hline
\end{tabular}
\end{center}
\end{table}

\section{Variation of aperture with roughness model}
\label{roughness}

Current models of lunar surface roughness use a single surface
slope over the entire length of the cascade through which
radiation refracts. This is only an approximate treatment, as
discussed by James \& Protheroe \cite{JamesProtheroeLCT109},
since it takes the typical deviation at scales of order a
wavelength, and treats it as a `large-scale' (greater than a
cascade length) phenomena. Here we define `small-scale'
roughness to be that on a scale between a cascade length
(typically a few metres for hadronic cascades) and the wavelength
in the medium. The consequences of restricting scales to this range is
discussed later.  The effect of small-scale roughness is
to reduce the coherence between radiation from different parts of
the cascade.  This in turn broadens the angular width over
which the radiation is emitted while reducing the peak strength,
and also allows transmission for angles of incidence greater
than the critical angle, which is an effect observed for rough
optical surfaces \cite{Griswold07}.

We carry out simulations using a toy model for an extreme
case in which radiation from different parts of a shower in a
near-surface cascade will see different surfaces and thus be
refracted semi-independently.  We refer to the resulting
effective aperture including our small-scale surface roughness model as
$A_{\rm eff}^R$, while the effective aperture for the ``standard
case'' without small-scale surface roughness is $A_{\rm eff}^S$. 
Because of interference effects, the
true effective aperture is likely to be between these two
extremes, and it may be reasonable to approximate this over a
restricted frequency range by an ``adjusted'' effective aperture
$A_{\rm eff}^A = r A_{\rm eff}^S + (1-r) A_{\rm eff}^R$.  First
we describe the toy model, then the simulation method and
the resulting aperture $A_{\rm eff}^R$. Interference effects and
a method of determining $r$ and hence $A_{\rm eff}^A$ are
dealt with, and
finally we discuss the validity of the approximations made and the necessity of
a rigorous treatment of the effect of small-scale roughness.

\subsection{Toy model calculations and results}

Assuming the smallest roughness scale to which radiation is sensitive is that of a wavelength,
a cascade of length $L$ might see up to $N_S$ ($\approx L_S/\lambda$)
refractive surface elements, where $L_S$ is the cascade length. Though it is unclear whether the applicable
wavelength $\lambda$ is that of the low-index or high-index medium, since our goal is to put an inclusive bound on the effects
of such roughness, we choose $\lambda=\lambda_n$, where $\lambda_n$ is the wavelength in the medium of 
highest refractive index. We also choose $\lambda$ corresponding to the frequency $\bar{f} = (f_{\rm min} \times f_{\rm max})^{0.5}$.
For our experiment at ATCA, $\bar{f}= 1.47$~GHz, so $\lambda_n=11.8$~cm in the regolith ($n=1.73$).

We calculate $N_S$ by scaling the energy- and medium-dependent $L_S$ from its value of
$12 \chi_0/\rho $ ($12$ radiation lengths given the medium density: $\sim 4.7$~m) for hadronic
cascades in ice at $10$~EeV \cite{Alvarez-Muniz97} by $7.5$\% per
decade in energy as per Williams \cite{Williams04}.
Since in our treatment, each portion of the cascade would `see' a different refractive element,
we break the cascade into $N_S$ separate segments, rounding $N_S$ up to take an integer value.
Thus we arrive at a shower length given by Eq.\ \ref{eq_shower_length},
and the number of shower segments $N_S$ given by Eq.\ \ref{eq_shower_segments}:
\begin{eqnarray}
L_S & = & 12 (X_0/\rho) \left[1+0.075 \, \log_{10} \left(\frac{E_S}{10~{\rm EeV}} \right) \right] \label{eq_shower_length} \\
N_S & = & L_S \left( \frac{n \bar{f}}{c} \right) \label{eq_shower_segments}
\end{eqnarray}
We assume that each of the $N_S$ segments contains an equal portion of the total excess
tracklength of each cascade, so that the peak electric field amplitude from each is $1/N_S$ that of the
cascade as a whole. We also assume that the segment has length $L_S/N_S$, so the width of the
Cherenkov cone is correspondingly broadened by the same factor $N_S$. Table
\ref{subshower_table} summarises the scaling relationships and number of shower segments for
hadronic cascades at energies of $10^{21}$, $10^{22}$, and $10^{23}$~eV in the regolith and
megaregolith.

The radiation from each
cascade segment is treated independently, with the simulated experiment able to detect
none, all, or some of the cascade segments, with the detection probability of the primary
neutrino being equal to one minus the probability that none of the individual
cascade segments are detected. We calculate the transmission coefficients separately
for each piece of surface, which results in the observed
signal appearing to come mostly from those parts of the surface pointing
roughly towards the observer. This method is identical to that described by James
and Protheroe \cite{JamesProtheroeLCT109} for handling cascades from interactions of secondary
$\mu$, $\tau$, and $\nu$. However, it ignores the possibility of interference between the
radiation from cascade segments, an effect which is unimportant for secondary interactions
separated by large distances. We make an approximate adjustment for such interference by calculating an
`adjusted' effective aperture $A_{\rm eff}^A$ in Sec.\ \ref{interference} -- suffice to say
for now that $A_{\rm eff}^A$ must lie between $A_{\rm eff}^S$ and $A_{\rm eff}^R$.

\begin{table*}
\begin{center}
\caption{Parameters relating to the number of sub-showers in the surface roughness estimate, and some typical parameter values. Energies relate to the total hadronic cascade energy, which is typically $20$\% that of the primary neutrino energy. \label{subshower_table}}
\renewcommand{\arraystretch}{1.2}
\begin{tabular}{l | ccc | ccc | c |ccc}
\hline
\hline
& $X_0$ & $\rho$ & $X_0/\rho$ & \multicolumn{3}{c|}{$L_S$ (m)} & $\lambda_{\bar{f}}/n$ & \multicolumn{3}{c}{$N_s$} \\
Layer & (g/cm$^2$) & (g/cm$^3$) & (cm) & $10^{21}$~eV& $10^{22}$~eV& $10^{23}$~eV & (m) & $10^{21}$~eV& $10^{22}$~eV& $10^{23}$~eV \\
\hline
Regolith & 22.59 & 1.8 & 12.55 & 1.51 & 1.62 & 1.73 & 0.118 & 13 & 14 & 15 \\
Megaregolith & 22.59 & 3.0 & 7.53 & 0.895 & 0.962 & 1.03 & 0.082 & 11 & 12 & 13 \\
\hline
\hline
\end{tabular}
\end{center}
\end{table*}

To generate a surface roughness deviate, we first generate a deviate on the length scale $L_S$ of the entire cascade as per James \& Protheroe \cite{JamesProtheroeLCT109}, where the slope tangents in each direction are sampled from a normal distribution with mean $0$ and standard deviation $\tan S_{\rm rms}$ given according to Eq.\ \ref{surf_rough} (modified from Olhoeft \& Strangway \cite{Olhoeft75} by substituting $L_S/{\rm 1 m}$ for $\lambda/{\rm 1 cm}$):
\begin{eqnarray}
\tan S_{\rm rms} & = & 0.105 \, L_S^{-0.22} \label{surf_rough}
\end{eqnarray}
The distribution of surface roughness experienced by radiation at wavelength $\lambda_n$ (m) is found be replacing $L_S$ by $\lambda_n$ in the above equation. The extra roughness at small length scales can be thought of as additional small-scale deviation $\Delta \tan S_{\rm rms}$ superimposed over the large-scale roughness, which must produce the correct total value of $S_{\rm rms} (\lambda_n)$. Since the small-scale deviation in this model is independent of the large-scale deviation, we find:
\begin{eqnarray}
\tan^2 S_{\rm rms}(\lambda_n) & = & \tan^2 S_{\rm rms}(L_S) + (\Delta \tan S_{\rm rms})^2 \label{rms_addition}
\end{eqnarray}
Substituting the formula for $\tan S_{\rm rms}$ from Eq.\ \ref{surf_rough} into Eq.\ \ref{rms_addition} then gives the value of the additional surface deviate:
\begin{eqnarray}
\Delta \tan S_{\rm rms} & = &  0.105 \left( \lambda_n^{-0.44}-L_S^{-0.44} \right)^{0.5} \label{deltan}
\end{eqnarray}
This gives a typical value for $\Delta \tan S_{\rm rms}$ of $8^{\circ}$; we approximate $\Delta\tan S$ for each individual sub-showers to be independent of each other for simplicity. Thus, for each cascade, we first calculate the length and generate a bulk surface normal by randomly sampling two slope tangents according to Eq.\ \ref{surf_rough}, then for each sub-shower we modify the bulk surface normal by adding extra surface tangents sampled according to Eq.\ \ref{deltan}.

\begin{figure*}
\begin{center}
\includegraphics[width=0.8\textwidth]{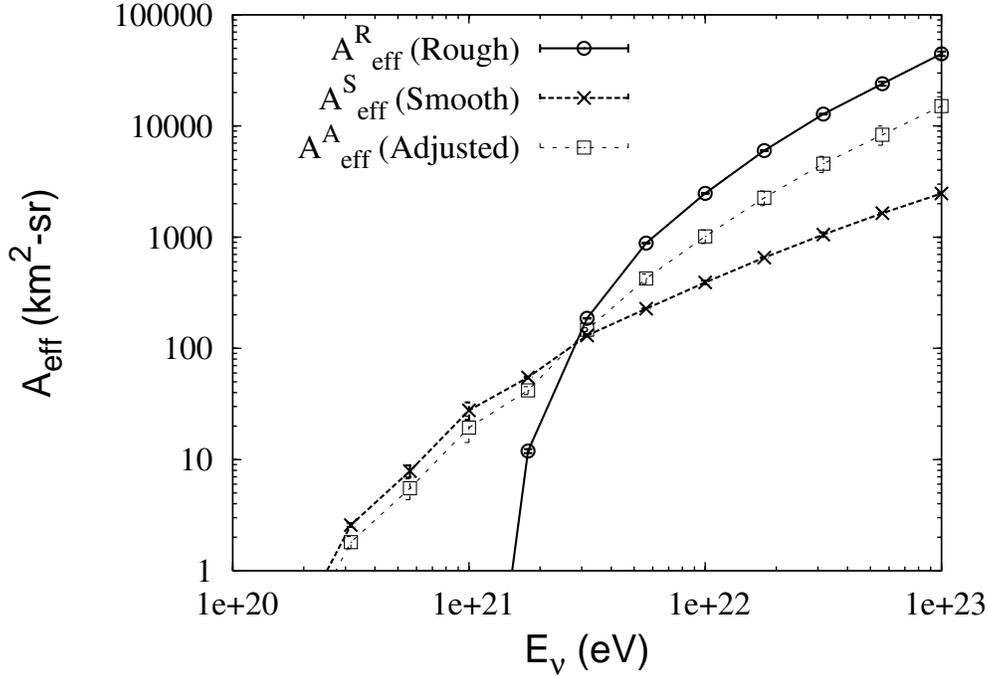}
\caption{Effective apertures of a fictitious experiment, calculated using large-scale roughness only ($A_{\rm eff}^S$),
our small-scale roughness approximation ($A_{\rm eff}^R$), and an adjusted aperture which is a linear
combination of them both ($A_{\rm eff}^A$ -- see Sec.\ \ref{interference}).
\label{aeffcomp}}
\end{center}
\end{figure*}

The additional complexity caused by breaking the cascade into segments means that we model only a simplified version
of our experiment, using three antennas with a flat bandpass and single, circularly-polarised receivers operating in
coincidence, with thresholds only approximately that of our real experiment. We run the simulation both with
(treatment described above) and without (standard treatment) small-scale surface roughness for primary
neutrino energies in the range $10^{20}<E_{\nu}<10^{23}$~eV. Though the absolute values of the effective apertures $A_{\rm eff}^S$
and $A_{\rm eff}^R$ have little meaning for this fictional experiment, they are still illustrative to plot, which we do in Fig.\ \ref{aeffcomp}.

Comparing $A_{\rm eff}^S$ with $A_{\rm eff}^R$, the effect of small-scale
roughness is significant. Since the peak emission from each cascade segment
is a factor of $N_S$ ($\sim 10$) lower than that of the whole cascade, the effective
neutrino energy detection threshold has been increased by a factor of the
same order. However, at the highest energies ($\gtrsim 10^{22}$~eV), the
probability of detection has increased by a factor of order $100$, since the
emission from each cascade is broader, and there are more cascades. In all,
the effect of small-scale roughness on the detection probability mimics
that of observing at a lower frequency in the case when no such roughness
is considered, though the expected time-domain signature would be quite different.

\subsection{Estimate of interference effects}
\label{interference}

The aperture $A_{\rm eff}^R$ calculated by modeling small-scale surface roughness,
as previously noted, excludes interference between
different cascade segments. Unlike radiation from two separate cascades, which if
exiting the Moon in the same direction would be seen as two independent signals by
a detector, the arrival times of radiation from two cascade segments will likely be
separated by less than their duration -- that is, they interfere. At one extreme (`case 1'),
all the cascade segments will see the same surface and their radiation will exit in
the same direction. Thus they interfere according to the standard Cherenkov condition,
and the standard modeling producing $A_{\rm eff}^S$ is the correct treatment. At
another extreme (`case 2'), the refracted radiation patterns from each cascade segment will not
overlap, the cascades can be treated independently, and the results generated from
our small-scale roughness model (i.e.\ $A_{\rm eff}^R$) will be correct insofar as
the surface generated is appropriate.

\begin{figure*}
\begin{center}
\includegraphics[width=0.8\textwidth]{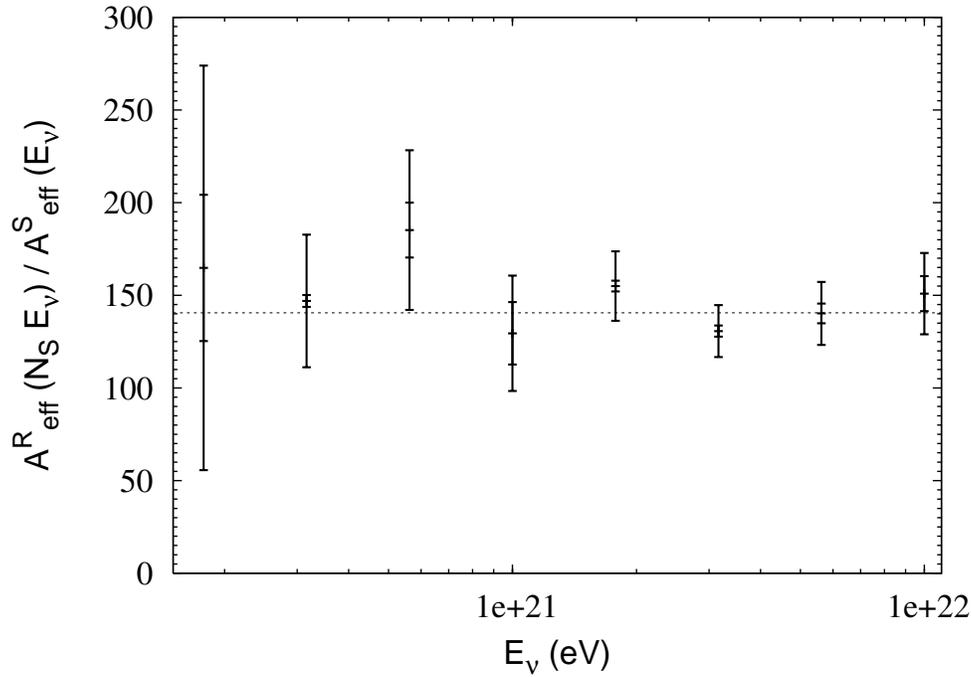}
\caption{The ratio $A_{\rm eff}^R (N_S \, E_{\nu})/A_{\rm eff}^S (E_{\nu})$. The error bars reflect
both uncertainties in the apertures themselves from the Monte Carlo simulation (inner error bars),
and also including the error from $N_S = 13 \pm 1$. The fit is to the range $E_\nu \ge 10^{21}$~eV only.
\label{aperture_ratios}}
\end{center}
\end{figure*}

We model our results as a linear combination of these two extremes by comparing
the calculated aperture $A_{\rm eff}^S (E_{\nu})$ to that of $A_{\rm eff}^R (N_S \, E_{\nu})$.
On average, the peak strength of the emission of each cascade in the $A_{\rm eff}^S (E_{\nu})$
calculation will be the same as that of each cascade segment in the $A_{\rm eff}^R (N_S \, E_{\nu})$
calculation, while the width of the radiation patterns from the segments under rough
modeling will be $k_L \, N_S$ times greater, where $k_L$ allows for the slow growth of
cascade length with energy:
\begin{eqnarray}
k_L & = & (1+0.075 \log_{10} N_S)^{-1}.
\end{eqnarray}
Assuming no interference, the aperture in the rough `R' case will be one factor
of $N_S$ times larger than the standard `S' case due to there being $N_S$ as many
sub-cascades, and a further factor of $k_L N_S$ due to the increased width as
described above. Thus we expect that the ratio
$A_{\rm eff}^R (N_S \, E_{\nu})/A_{\rm eff}^S (E_{\nu})$ should be
$k_L \, N_S^2 \, \sigma_H(N_S \, E_{\nu})/\sigma_H(E_{\nu})$,
where $\sigma_H$ is the cross-section for interactions producing hadronic cascades.
This assumes that the interaction rate is proportional to the cross-section, which is
accurate to order $10$\% (see Appendix \ref{xsectionapp}). We also
assume that all hadronic interactions are dominated by neutrino-nucleon interactions
(also an order $10\%$ approximation \cite{JamesProtheroeLCT109}), so that
$\sigma_H \propto E^{0.363}$ \cite{Gandhi98}, and thus:
\begin{eqnarray}
A_{\rm eff}^R (N_S \, E_{\nu})/A_{\rm eff}^S (E_{\nu}) & = & k_L N_S^{2.363}.
\end{eqnarray}
If the interference is complete (radiation always exits in an
identical direction), and we assume a detection probability of $1$ within some
part of the Cherenkov cone and $0$ otherwise so that there is no gain
in detection probability from seeing the emission from two cascade segments,
then only one of the $N_S$ cascades will be detected in the $A_{\rm eff}^R (N_S \, E_{\nu})$
simulation, and we should find the ratio:
\begin{eqnarray}
A_{\rm eff}^R (N_S \, E_{\nu})/A_{\rm eff}^S (E_{\nu}) & = & k_L N_S^{1.363}.
\end{eqnarray}
We thus define the fractional overlap $r$ such that $r=0$ indicates no interference
($A_{\rm eff}^R$ applies), and $r=1$ complete interference ($A_{\rm eff}^S$ applies),
i.e.:
\begin{eqnarray}
A_{\rm eff}^R(N_S \, E_{\nu})/A_{\rm eff}^S (E_{\nu}) = (1-r) k_L N_S^{2.363}+ r \, k_L N_S^{1.363}.
\end{eqnarray}
Using $N_S = 13 \pm 1$, we plot $A_{\rm eff}^R (N_S \, E_{\nu})/A_{\rm eff}^S (E_{\nu})$
in Fig.\ \ref{aperture_ratios}, and fit for $r$ to the high-energy
regime where $r$ is constant. Note that in this model the number $N_s$ of surface pieces refers 
to the number upon which the radiation is incident, not the number 
an observer sees. We expect an increasing ratio at low energies where a larger fraction
of the emitted radiation is partially detectable (detection probability being neither $0$ nor $1$),
since then there is a gain in
calculated aperture from two cascade segments radiating in the same direction. Though an
increasing value of $r$ is not observed at low energies -- likely the effect is obscured by the large uncertainties -- we nonetheless
use only the range $E \ge 10^{21}$~eV for the fit, finding $r=0.70 \pm 0.06$.

An adjusted aperture $A_{\rm eff}^A$ can then be
calculated by $A_{\rm eff}^A(E_{\nu}) = r A_{\rm eff}^S(E_{\nu}) + (1-r) A_{\rm eff}^R(E_{\nu})$. This has added to Fig.\
\ref{aeffcomp} for our fictitious experiment. By assuming similar behaviour for our real experiment with the ATCA,
we can calculate an adjusted aperture from our standard aperture only:
\begin{equation}
A_{\rm eff}^A (E_{\nu}) = 0.7 A_{\rm eff}^S (E_{\nu}) + 0.3 k_L N_S^{2.363} A_{\rm eff}^S (E_{\rm \nu}/N_S).
\end{equation}
The result has already been given for both our experimental
aperture and limit in Figs.\ \ref{ATCA_isotropic_app} and
\ref{ATCAisotropic_lim} respectively, by using the aforementioned
values of $r$ and $N_S$.

\subsection{Assessment of accuracy}
\label{accuracy}

The model for small-scale surface roughness presented here is
intended as a toy model which deals only with the most
significant possible effects of small-scale roughness at high
frequencies. For instance, the sudden change in slope around $2
\times 10^{21}$~eV in Fig.~\ref{aeffcomp} is purely an
artefact of our superposition of two extreme models --
there is no reason to expect that a rigorous calculation
would produce any energy-dependent results sharper than the slow
(logarithmic) increase in cascade length with particle
energy.

In cases where the surface slopes are positively correlated, and/or when a cascade is
sufficiently deep that the far-field conditions are satisfied at the surface over most of
the cascade, then our model of treating the radiation from each cascade/surface piece
independently breaks down. While our introduction of the overlap parameter `r' goes some way to
correcting for such correlation, in these cases the value of `r' would likely be larger,
so the effects of small-scale roughness estimated by our toy model are more likely to be over-estimates
than under-estimates.

Our model also ignores roughness on scales smaller than a
wavelength, which can not be dealt with by splitting the cascade
into segments as in our toy model (it is not possible to produce
multiple-refraction effects for structures of size less than that
of a wavelength). However, we do expect such roughness to broaden
the emission at the expense of peak refracted field strength --
for our current model, this would mostly cause an increase in the
overlap ratio $r$ in our results.  Therefore, we ignore
sub-wavelength roughness until a more complete model can be
constructed, which we leave to a future work. An additional
approximation is that we have assumed only a single
frequency for our surface-roughness estimates, rather than a
continuous range.

Despite these shortcoming, we have developed the first treatment
of roughness on scales smaller than a cascade length, an
important effect for high-frequency observations that has been
ignored in all previous calculations. Whether this effect helps
or hinders neutrino detection will depend upon the shape of the
UHE neutrino spectrum. What we can say is that the difference in detection probability
between high- and low-frequency observations may not be as
dramatic as previously thought, but that a reconstruction of
cascade parameters from a detected event may prove more
difficult.


\begin{acknowledgments}
The Australia Telescope Compact Array is part of the Australia
Telescope which is funded by the Commonwealth of Australia for
operation as a National Facility managed by CSIRO. This research
was supported by the Australian Research Council's Discovery
Project funding scheme (project numbers DP0559991 and
DP0881006). J.A-M thanks Xunta de Galicia (PGIDIT 06 PXIB 206184
PR) and Conseller\'\i a de Educaci\'on (Grupos de Referencia
Competitivos -- Consolider Xunta de Galicia 2006/51).
\end{acknowledgments}

\end{document}